\documentstyle[12pt,epsf]{article}

\title{An Effective Equation of State for Dense Matter with Strangeness}
\author{
Shmuel Balberg and Avraham Gal\\ {\sl \small The Racah Institute of
Physics}\\ {\sl \small The Hebrew University, Jerusalem 91904,
Israel}\\ }
\date{ }
\begin{document}
\setcounter{page}{1}

\maketitle
\begin{abstract}
An effective equation of state which generalizes the
Lattimer-Swesty equation for nuclear matter is presented for
matter at supernuclear densities including strange baryons.
It contains an adjustable baryon potential energy density,
based on models of local potentials for the baryon-baryon
interactions. The features of the equation rely on the
properties of nuclei for the nucleon-nucleon interactions,
and mainly on experimental data from hypernuclei for the
hyperon-nucleon and hyperon-hyperon interactions. The
equation is used to calculate equilibrium compositions and
thermodynamic properties of high density matter with
strangeness in two astrophysical contexts: neutron star
matter (transparent to neutrinos) and proto-neutron star
matter (opaque to neutrinos). The effective equation of state 
reproduces typical properties of high density matter found in 
theoretical microscopic models. Of these, the main result 
is that hyperons appear in both types of matter at about twice 
the nuclear saturation density, and that their appearance 
significantly softens the equation of state. The range of maximal 
masses of neutron stars found in a comprehensive parameter survey is 
1.4-1.7 $M_\odot$. Another typical result is that the maximal 
mass of a proto-neutron star with strange baryons is higher than 
that of an evolved neutron star (opposite to the case of nuclear matter), 
setting the stage for a ``delayed collapse" scenario.
\end{abstract}

\setlength{\parskip}{0.0in}


\setlength{\textheight}{9.0in}
\setlength{\parskip}{0.1in}
\newpage


\section{Introduction}

\hspace* {6mm}

The properties of matter at supernuclear densities are a key subject of 
interest in the study of core-collapse supernovae and neutron stars. Theories 
of such matter typically predict that at the nuclear saturation density, 
$\rho_{s}\!=\!0.15-0.16$ fm$^{-3}$, it is composed of free nucleons and 
leptons, but at higher densities several more species of particles may appear. 
Chiefly among these are strange baryons, namely, the $\Lambda$, $\Sigma$ and 
$\Xi$ hyperons, while $\Delta$ baryons along with pion and kaon condensations 
and a deconfined-quark-phase have also been considered. The collapsed core in 
type II supernovae and neutron stars serve, in this sense, as cosmological 
laboratories for hadronic physics.

The appearance and effects of such ``novel'' particles in dense matter have 
been included in several studies in the context of neutron star structure 
(see for example, Refs. [1-5]). Some work has also been done regarding the 
collapsing core in type II supernovae \cite{MayTaWi}, and recently attention has 
also been given to the possible impact of the appearance of strange particles in 
proto-neutron star evolution [7-10].

The physical state and composition of matter at high densities depends mainly 
on the nature of the strong interactions. However, the theoretical details of these 
interactions are not completely understood, and uncertainties also 
arise from the incompleteness of the available experimental data for nuclear 
matter. Such uncertainties are compounded when strange hadrons are taken into 
account, as relevant experimental data is rather scarce. Theoretical studies of 
nuclear and hadronic matter are usually performed with sophisticated models such 
as relativistic mean field (RMF) and Brueckner-Hartree-Fock, which involve 
effective Lagrangians and/or specific assumptions about baryon-baryon 
interactions. Some applications of these models are technically complex and often 
prove expensive in terms of CPU time when performing calculations at various 
combinations of densities and temperatures. Thus, there is also an interest in 
using an effective equation of state (EoS), which is analytic and depends on a 
small number of parameters. Such an equation will be convenient as a supplementary 
tool, mainly for direct use in rapid hydrodynamical simulations, and for 
conducting extensive parameter surveys that are necessary given the large 
uncertainties in the theory and experimental data regarding strong 
interactions.

Two commonly used effective equations of state for 
hot (finite temperature) nuclear matter (nucleons and leptons) are those of 
Lattimer and Swesty \cite{LS} and Baron, Cooperstein and Kahana \cite{BCK}. In 
the present study we wish to construct a generalized effective equation of state 
for high density hadronic matter, by extending the Lattimer-Swesty (hereafter 
LS) equation so as to include additional species of particles. We currently 
limit ourselves to strange baryons, which we view as the prime candidates 
for appearing in high density matter.

In Section 2 we discuss our approach for constructing the effective EoS, 
basing it on the formulation of local (effective) density dependent potentials
 for the various cases of the baryon-baryon interactions (lepton interactions 
are assumed to be negligible, and thus leptons are treated as non-interacting 
particles). Section 3 outlines the equilibrium conditions, which determine 
the composition and physical properties of high density matter for 
neutrino-free matter (neutron star) and neutrino trapped matter (core-collapse 
and proto-neutron-star) in beta equilibrium.

Our choice of parameters for these equations, derived mainly on the basis of 
hypernuclear data, is presented in Section 4; we note, that these parameters 
allow accurate modeling of field-theoretical interactions. 
We also specifically treat the consequences of recent theoretical and 
experimental work on $\Sigma^{-}$-atoms \cite{BattFrGal,MarFrGalJen}, which seem to 
suggest a strong isoscalar repulsion in the interaction of $\Sigma$ hyperons with a 
bulk of nucleons, unlike the commonly assumed interaction.

Results for the equilibrium compositions of high density matter and the 
corresponding EoS are presented in Section 5. These EoS's are 
examined with regard to the static neutron star masses they predict in Section
 6. In Section 7 we offer conclusions and further discussion.

\section{The Effective Equation of State}
The construction of an effective EoS for supernuclear density matter centers 
on the description of the strong interaction energy. In this approach one 
models these interactions in some analytic form, involving a small number of 
parameters.

The internal energy density is approximated by dividing it 
into a baryon contribution, which includes kinetic, potential and mass terms, 
and a lepton contribution which has only kinetic and mass terms. This 
approximation, introduced in the LS EoS, yields the following structure of the 
internal energy density for supernuclear dense matter:
\begin{equation} \label{eq:geneps}
  \varepsilon_{tot}=\varepsilon_{kin}(\rho,T,\{x_{i}\})+
    \varepsilon_{pot}(\rho,\{x_{i}\})+\varepsilon_{mass}(\rho,\{x_{i}\})+
    \varepsilon_{kin}(\rho,T,\{l_{i}\})+\varepsilon_{mass}(\rho,\{l_{i}\})\;   ,
\end{equation}
where $\rho$ is the total baryon density, $T$ is the matter temperature (whose
 effect is handled in the kinetic energy terms), $\{x_{i}\}$ and $\{l_{i}\}$ 
are respectively the baryon 
and lepton fractions, $x_{i}=\rho_{x_{i}}/\rho; l_{i}=\rho_{l_{i}}/\rho$. For 
finite temperatures, a photon contribution must be added as well.

The separation of the lepton contribution is made possible by treating leptons 
as non-interacting (relativistic) particles, and, as discussed in Ref. 
\cite{LatPethRavLam}, electron screening effects may be ignored because the 
electron-screening length is larger than the baryon separation distance. 

The heart of the effective EoS is, of course, casting the potential energy 
density term, $\varepsilon_{pot}$, into an effective form. This is achieved by 
assuming local density dependent potentials for all strong interactions. 
Such a potential model must reproduce the basic features of the strong 
interactions, i.e. long-range attraction and short range repulsion, and - in 
some cases - charge dependence, compatible with isospin invariance. The most 
economic way to construct such a potential for the case of a single baryon of 
species $y$ in bulk matter of baryon species $x$ of number density $\rho_{x}$ 
is:
\begin{equation} \label{eq:genlocpot}
   V_y (\rho_{x}) = a_{xy}\rho_x + 
                    b_{xy}t_x t_y \rho_{x} + 
                    c_{xy}\rho_x^{\gamma_{xy}}\;    .
\end{equation}

The first term yields attraction ($a_{xy}$ negative), the third term yields 
repulsion ($c_{xy}$ positive, supposedly introducing multibody interactions), 
and the second (symmetry) term introduces the charge dependence through a charge 
(isospin) $t$. The common choice of parameters in effective approaches \cite{LS,BCK} 
favors a linear density dependence of the attraction and symmetry terms, while 
$\gamma_{xy}$ must be greater than unity, so that repulsion will 
dominate at high densities (short ranges). This formulation limits the potential 
energy model to at most four parameters, whose values are chosen to reproduce 
experimental data and accepted theoretical results. Eq. (\ref{eq:genlocpot}) 
can, of course, also be used for the case of a single species (i.e., $y\!=\!x$), 
describing the local potential felt by each baryon in the bulk of baryons of the 
same species.

We note that a local effective potential for the nucleon-nucleon interaction 
provides the basis for the well known Skyrme model (see Ref. \cite{Vauth} for a recent 
review). Similar modeling for hyperons in nuclear matter was first performed by 
Millener, Dover and Gal (\cite{MDG}, hereafter MDG), on the basis of experimental data 
of various $\Lambda$-hypernuclei, deriving a local potential for a single $\Lambda$ 
in nuclear matter.

The local potential for a single baryon in a bulk of other baryons may be 
extrapolated into the potential energy density of bulk matter with a total density 
$\rho$ that includes both types of baryons. This is done by folding $V_y(\rho_x)$ 
with the partial density $\rho_y$, and vice versa ($V_x(\rho_y)$ with $\rho_x$), 
combined with weight factors (avoiding double counting of each interaction). If we 
choose the weight factors to be $x/(x+y)$ and $y/(x+y)$ respectively, the 
contribution of the interactions between baryons of species $x$ and baryons of 
species $y$ to the internal energy density, $\varepsilon(x\leftrightarrow y,\rho)$ 
will be:
\begin{equation} \label{eq:epotxy}
   \varepsilon_{pot}(x\leftrightarrow y,\rho)=a_{xy}xy\rho^2+
                                     b_{xy}t_xt_y xy\rho^2+
                                     c_{xy}\left(\frac{x}{x+y}x^{\gamma_{xy}} y+
                                     \frac{y}{x+y}y^{\gamma_{xy}}x \right)
                                               \rho^{\gamma_{xy}+1}\;    ,
\end{equation}
where $x$ and $y$ are defined as $x\!=\!\rho_{x}/\rho$, $y\!=\!\rho_{y}/\rho$.

The above weight factors do not appear in the first two terms as they just add to 
unity, while in the repulsion term they retain discrimination between baryons $y$ in 
matter of baryons $x$ and baryons $x$ in matter of baryons $y$. Although this form is 
not unique, the weight factors $x/(x+y)$ and $y/(x+y)$ yield the expected results 
in the case $x\!=\!y$ and for the limits $x\!>>\!y$ and vice versa. Inherent in 
Eq. (\ref{eq:epotxy}) is the neglect of possible screening effects of other 
baryons, so that the interaction between any two baryons is dependent only on the 
distance between them. This interaction may be taken as an effective average 
interaction between any two baryon species.

For the specific case of baryons of a single species, setting $x=y$ in Eq. 
(\ref{eq:epotxy}) requires a factor of $\frac{1}{2}$  to avoid double counting:
\begin{equation} \label{eq:epotxx}
   \varepsilon_{pot}(x\leftrightarrow x,\rho)=\frac{1}{2}\left[a_{xx} x^2\rho^2+
                                        b_{xx}t_x^2 x^2\rho^2+
                                        c_{xx}x^{\gamma_{xx}+1}\rho^{\gamma_{xx}+1}
                                                   \right] \;    .
\end{equation}
For nucleons, charge independence is respected by choosing  $a_{nn}=a_{pp}=a_{np}
\equiv a_{NN}$, with similar relations among the symmetry and repulsion 
coefficients. Thus, neutrons and protons can be treated as members of a single species 
with the appropriate symmetry term, and we can then retrieve the potential energy 
density of nucleons, similar to the form presented by LS (who denote the power in 
the repulsion term by $\delta$):
\begin{equation} \label{eq:epotNN}
   \varepsilon_{pot}(n,z,\rho)=\frac{1}{2}\left[a_{NN}(n+z)^2\rho^2+
                                          b_{NN}(n-z)^2\rho^2+
                                          c_{NN}(n+z)^{\delta+1}\rho^{\delta+1}
                                                   \right] \;    .
\end{equation}
LS use a slightly different expression, $a'_{NN}(n+z)^2\rho^2+b'_{NN}nz\rho^2+
c_{NN}(n+z)^{\delta+1}\rho^{\delta+1}$, which requires a rearrangement of the 
coefficients in Eq. (\ref{eq:epotNN}). In reaching Eq. (\ref{eq:epotNN}), 
$(n-z)$ repulsive contributions were dropped so that the short-range 
repulsion term depends only on the total nucleon density.

We now turn to high density matter, where the hyperon species $\Lambda$, $\Sigma$ 
and $\Xi$ are considered in addition to the nucleons. The full potential energy 
density expression will include several new terms, for the different baryon-baryon 
combinations:
\begin{eqnarray} \label{eq:epotbar}
   \varepsilon_{pot}(\rho) = \!\!& &\!\!
   \frac{1}{2}\left[a_{NN}N^2\rho^2+b_{NN}(n-z)^2\rho^2+
                     c_{NN}N^{\delta+1}\rho^{\delta+1}\right]  \\ 
   \!\! & & \!\!\!\!+\:a_{\Lambda N}N\Lambda\rho^2+
       c_{\Lambda N}\left(\frac{N}{N+\Lambda}N^\gamma\Lambda+
       \frac{\Lambda}{N+\Lambda}\Lambda^\gamma N\right)\rho^{\gamma+1} 
                                                                 \nonumber \\
   \!\! & & \!\!\!\!+\:\frac{1}{2}\left[a_{\Lambda\Lambda}\Lambda^2\rho^2
                    +c_{\Lambda\Lambda}\Lambda^{\gamma+1}\rho^{\gamma+1}
   +a_{\Xi\Xi}\Xi^2\rho^2+b_{\Xi\Xi}(\Xi^--\Xi^0)^2\rho^2+
            c_{\Xi\Xi}\Xi^{\gamma+1}\rho^{\gamma+1}\right]     \nonumber \\
   \!\! & & \!\!\!\!+\:a_{\Xi N}N\Xi\rho^2+b_{\Xi N}(n\!-\!z)(\Xi^-\!-\!\Xi^0)\rho^2+
       c_{\Xi N}\left(\frac{N}{N+\Xi}N^\gamma\Xi+
       \frac{\Xi}{N+\Xi}\Xi^\gamma N\right)\!\rho^{\gamma+1}
                                                                 \nonumber \\
   \!\! & & \!\!\!\!+\:a_{\Lambda \Xi}\Xi\Lambda\rho^2+
       c_{\Lambda \Xi}\left(\frac{\Lambda}{\Xi+\Lambda}\Lambda^\gamma\Xi+
       \frac{\Xi}{\Xi+\Lambda}\Xi^\gamma \Lambda\right)\rho^{\gamma+1} 
                                                                 \nonumber \\
   \!\! & & \!\!\!\!+\:a_{\Sigma N}N\Sigma\rho^2+b_{\Sigma N}(n\!-\!z)(\Sigma^-\!-
       \!\Sigma^+)\rho^2+c_{\Sigma N}\!\left(\frac{N}{N+\Sigma}N^\gamma\Sigma+
       \frac{\Sigma}{N+\Sigma}\Sigma^\gamma N\right)\!\rho^{\gamma+1} 
                                                                 \nonumber \\
   \!\! & & \!\!\!\!+\:a_{\Lambda \Sigma}\Sigma\Lambda\rho^2+
       c_{\Lambda\Sigma}\left(\frac{\Sigma}{\Sigma+\Lambda}\Sigma^\gamma\Lambda+
       \frac{\Lambda}{\Sigma+\Lambda}\Lambda^\gamma\Sigma\right)\rho^{\gamma+1} 
                                                                 \nonumber \\
   \!\! & & \!\!\!\!+\:a_{\Sigma \Xi}\Sigma\Xi\rho^2+
              b_{\Sigma \Xi}(\Xi^-\!-\!\Xi^0)(\Sigma^-\!-\!\Sigma^+)\rho^2+
       c_{\Sigma \Xi}\!\left(\frac{\Xi}{\Xi+\Sigma}\Xi^\gamma\Sigma+
       \frac{\Sigma}{\Xi+\Sigma}\Sigma^\gamma \Xi\right)\!\rho^{\gamma+1}
                                                                 \nonumber \\
   \!\!& & \!\!\!\!+\frac{1}{2}\left[a_{\Sigma\Sigma}\Sigma^2\rho^2+
                        b_{\Sigma\Sigma}(\Sigma^--\Sigma^+)^2\rho^2+
       c_{\Sigma\Sigma}\Sigma^{\gamma+1}\rho^{\gamma+1}\right]   \nonumber  \;,
\end{eqnarray}
where we define for each baryon species $x=\rho_{x}/\rho$, and introduce a 
shortened notation of $N\!=\!n+z$; $\Xi\!=\!\Xi^0+\Xi^-$; $\Sigma\!=\!\Sigma^++\Sigma^0+
\Sigma^-$. There are no standard (isovector) symmetry terms for the $\Lambda$ and 
$\Sigma^0$ baryons, since they have zero isospin; a nonstandard (isotensor) 
symmetry term for $\Sigma$'s is disregarded. We denote the exponent of the 
density-dependent repulsion term for the nucleon-nucleon interaction by $\delta$, 
following LS, while all other repulsive interactions are assumed to have a common 
exponent $\gamma$, although further diversity is, of course, possible.

The kinetic energy density of the baryons contains the non-relativistic fermion 
kinetic energies:
\begin{equation} \label{eq:ekin}
   \varepsilon_{kin}(\{x_i\},\rho,T)=
                      \sum_{i}\frac{\hbar^2\tau_i(x_i,\rho,T)}{2m^*_i}\;    ,
\end{equation}
where $m^*_i$ is the density dependent effective mass of the baryons of 
species $i$ (with $c\!=\!1$). The entire temperature dependence of the baryon 
thermodynamic functions is implicitly included in this term through the Fermi 
functions
\begin{equation} \label{eq:taui}
   \tau_i(x_i,\rho,T)=\frac{1}{2\pi^2}
                      \left[\frac{2m^*_iT}{\hbar^2}\right]^{5/2}
                      F_{3/2}\left(\eta(x_i,\rho,T)\right)
\end{equation}
where the temperature is in units of energy (setting $k_B=1$). $F_{3/2}$ is 
the corresponding Fermi integral, generally defined as
\begin{equation} \label{eq:Fermfunc}
   F_{k}(\eta)=\int^\infty_0 u^k\left(1+e^{u-\eta}\right)^{-1}du \;    .
\end{equation}
The parameter $\eta_i$ is directly related to the temperature and density 
through the inverse relation
\begin{equation} \label{eq:etai}
   \eta_i(x_i,\rho,T)=
   F^{-1}_{1/2}\left[2\pi^2x_i\rho\left(\hbar^2/2m^*_iT\right)^{3/2}\right]\;    .
\end{equation}
In the case of zero temperature matter the Fermi function reduces to the simple 
expression $\tau_i=\frac{3}{5}\left(3\pi^2x_i\rho\right)^{2/3}x_i\rho$. Typical 
nuclear models predict that in the depth of bulk matter the effective mass of 
baryons, $m^*_i$, will be lower than the bare mass $m_i$. The effective masses 
may be adequately represented in the Skyrme fashion
\begin{equation} \label{eq:meffi}
                1/m^*_i=1/m_i+\beta_i\rho\;    .
\end{equation}                                                      
The scaling parameters $\beta_i$ arise, however, from the model of the baryonic 
forces, and cannot be assumed independently. The effective equation has no means 
to fully relate consistent effective masses and forces (unlike RMF models where 
effective masses result naturally), and this point must be considered separately.

Finally, the contribution of the baryonic mass density is obviously given by
\begin{equation} \label{eq:emass}
   \varepsilon_{mass}(\{x_i\},\rho)=
                      \sum_{i}x_i\rho m^*_i\;    .
\end{equation}              
The chemical potentials of the baryons are also immediately related to the 
parameter $\eta_i$ :
\begin{equation} \label{eq:mu_i}
\mu_i=\left.\frac{\partial(\varepsilon_{tot}/\rho)}{\partial x_i}
                                                   \right|_{\{\tau_i\},\rho}
         =\left.\eta_iT+\frac{\partial((\varepsilon_{pot}+\varepsilon_{mass})/\rho)}
          {\partial x_i}\right|_{\{\tau_i\},\rho} \;     .
\end{equation}   
In the special case of zero temperature matter $\eta_i$ is undefined, and the 
chemical potentials are found by analytically differentiating $\partial(\varepsilon_{tot}/\rho)\partial x_i$.

Lepton contributions to the energy density are calculated through the complete 
(relativistic) Fermi integrals. For sufficiently high temperatures ($T\!\geq\!1$ MeV), 
the thermodynamic functions for electrons and neutrinos (when assumed to be 
trapped in the matter) can be approximated by the ultra-relativistic limit. Keeping 
the lowest order mass corrections results in the following equations for the 
chemical potentials \cite{LS}:
\begin{equation} \label{eq:mu_leptons}
   \rho Y_l=\frac{g_l}{6\pi^2}\left(\frac{\mu_l}{\hbar}\right)^3
            \left[1+\mu^{-2}_l\left((\pi T)^2-\frac{3}{2}m^2_l\right)\right]
\end{equation}
where $m_l$ is the lepton rest mass (again in units of energy), 
$g_l$ is the spin degeneracy and $Y_l$ is the net lepton fraction per baryon 
(the number of particles minus the number of anti-particles). The chemical 
potential is then solved by
\begin{eqnarray} \label{eq:expmu_leptons}
                       &       \mu_l=r-q/r  \; ; & \\
     r=\left[\sqrt{q^3+t^2}+t\right]^{1/3}\; ; & \;
     t=3\pi^2\hbar^3\rho Y_l/g_l\;  ;  \;\;\; \mbox{and} &  
     q=\frac{1}{3}\pi^2T^2-\frac{1}{2}m^2_l \nonumber \;    .
\end{eqnarray}

The contribution of the ultra-relativistic leptons to the total energy density for 
each lepton species $l$ is given by:
\begin{equation} \label{eq:e_leptons}
   \varepsilon_l=\frac{g_l\mu_l}{8\pi^2}\left(\frac{\mu_l}{\hbar}\right)^3
    \left[1+\mu^{-2}_l\left(2\pi^2T^2-m^2_l\right)+
    \frac{\pi^2T^2}{\mu^4_l}
    \left(\frac{7}{15}\pi^2T^2-\frac{1}{2}m^2_l\right)\right] \;    .
\end{equation}
Obviusly, in zero temperature matter all formuals for the lepton thermodynamic 
functions reduce to simple zero-temperature fermion gas functions.

We note that the muon cannot be treated as ultra-relativistic, and its thermodynamical 
functions must be calculated through the corresponding Fermi integrals. However, in the 
case of neutrino-trapped matter mouns are almost extinct (unlike the neutrino-free 
matter case, where muons form at significant fractions), and we shall disregard 
them for the case of finite temperature matter.

The construction of the equation of state is completed by the calculation of the 
various contributions to the pressure in the matter. The pressure is derived 
from the total energy density by:
\begin{equation} \label{eq:p_gen}
   P=\sum_i{x_i\rho\mu_i}+\sum_l{Y_l\rho\mu_l}-f
   =\sum_i{x_i\rho\mu_i}+\sum_l{Y_l\rho\mu_l}+Ts-\varepsilon\;    ,
\end{equation}
where $f$ is the free energy density, $f=\varepsilon-Ts$, and $s$ is the entropy 
density. Again, the pressure is composed of a baryon contribution and a lepton 
contribution, easy to derive from Eqs. (\ref{eq:epotbar})-(\ref{eq:mu_i}) 
and (\ref{eq:mu_leptons})-(\ref{eq:e_leptons}) respectively. In finite temprature 
matter the entropy per baryon (the commonly used variable in hydrodynamical 
simulations), $\tilde{s}$, for each baryon species is
\begin{equation} \label{eq:s_perb}
     \tilde{s}_i=\frac{1}{\rho}\left(\frac{5\hbar^2\tau_i}{6m^*_iT}
                                                -x_i\rho\eta_i \right)\;    ,
\end{equation}
and for each of ultra-relativistic lepton species is
\begin{equation} \label{eq:s_perl}
  \tilde{s}_l=\frac{1}{\rho}\frac{g_lT\mu^2_l}{6\hbar^3}
   \left[1+\mu^{-2}_l\left(\frac{7}{15}\pi^2T^2-\frac{1}{2}m^2_l\right)\right]\;   .
\end{equation}

Finite temperature calculations must include a photon gas $\mu_\gamma=0$ 
contribution, with:
\begin{equation} \label{eq:phot_cont}
  \varepsilon_\gamma=\frac{\pi^2T^4}{15\hbar^3}\;   ;
  P_\gamma=\frac{1}{3}\varepsilon_\gamma \;\;\;
  \mbox{and}\;\;\; \tilde{s}_\gamma=\frac{1}{\rho}\frac{4P_\gamma}{T}\;   .
\end{equation}

\section{Equilibrium Conditions}

The composition of the high density matter is obviously constrained by charge 
neutrality and baryon number conservation. Further constraints are imposed by 
the assumption of complete beta-equilibrium as the time scales of the weak 
interactions are generally much shorter than dynamical time scales in the 
relevant stellar systems. It is this constraint that sets most of the 
equilibrium conditions, as it forces all processes to take place in absolute 
chemical equilibrium.

The balance between baryons and leptons is maintained through beta decays. 
In nuclear matter, this is simply the equilibrium of the process 
$p+e^-\leftrightarrow n+\nu_e$. This equilibrium establishes the relation
\begin{equation} \label{eq:nu_eqlbm}
  \mu_p+\mu_e=\mu_n+\mu_{\nu_e}\; .
\end{equation}
Strange baryons may form in the matter through various types of weak interactions. 
For example, the $\Lambda$ hyperon can appear both in leptonic processes, 
$p+e^-\rightarrow\Lambda+\nu_e$, and in baryonic processes such as 
$n+n\rightarrow\Lambda+n$ (the inverse of which occurs in hypernuclei). There 
are also other possible processes, involving pions and kaons; however in this 
study we assume that meson condensates do not form in the matter (see discussion 
later in this section) and are therefore excluded. Note that the balance among 
strange baryons is maintained mainly through strong interactions like 
$\Lambda+n\rightarrow\Sigma^-+p$, or $\Lambda+\Lambda\rightarrow\Xi^-+p$. The 
resulting equilibrium conditions are simply that the chemical potentials of all 
baryons depend only on their electric charges (that determine the possible 
combinations in the various reactions):
\begin{eqnarray} \label{eq:bar_eqlbm}
  \mu_n=\mu_\Lambda=\mu_{\Sigma^0}=\mu_{\Xi^0}\; ; &
  \mu_p=\mu_{\Sigma^+}\; ; &
  \mu_{\Sigma^-}=\mu_{\Xi^-}=2\mu_n-\mu_p\; . 
\end{eqnarray}
It is noteworthy that the chemical potentials of the negatively charged hyperons 
are conditioned by a larger value than the neutral ones (since dense matter is 
typically dominated by neutrons). For example, studies of cold high density matter 
usually find that the first hyperon to appear is the $\Sigma^-$, closely preceding the 
lower-mass $\Lambda$ due to the charge difference.

The last fermion players in the high density matter are the muon and the muon neutrino. 
When the electron chemical potential exceeds the muon mass, muons and muon 
neutrinos can be formed by $e^-\rightarrow\mu^-+\nu_e+\bar{\nu}_{\mu}$, so that the 
equilibrium condition for the muon chemical potential is:
\begin{equation} \label{eq:emu_eqlbm}
  \mu_e=\mu_\mu+\mu_{\nu_e}+\mu_{\bar{\nu}_{\mu}} \; .
\end{equation}
Eqs. (\ref{eq:nu_eqlbm})-(\ref{eq:emu_eqlbm}) impose eight conditions on 
twelve variables, so that another four conditions are needed to uniquely determine 
the complete equilibrium composition of matter at given density and temperature. 
Two further conditions arise from charge neutrality and baryon number conservation. 
These conditions constrain the species fractions $\{x_i\}$ and $\{Y_j\}$ by 
\begin{equation} \label{eq:q=0}
  \sum_i{x_iq_i}+\sum_j{Y_jq_j}=0 \;
\end{equation}
where $\{q_i\}$ are the electric charges, and
\begin{equation} \label{eq:totx=1}
  \sum_i{x_i}=1 \;  .
\end{equation}
Finally, two more conditions result from the physical assumptions about the 
neutrino diffusion time scale. Core-collapse and proto-neutron star matter may be 
assumed to be opaque to neutrinos, so that neutrinos are trapped and the total 
lepton fraction must be kept constant. As the trapping in supernova occurs when 
the collapsing core reaches densities of about $10^{12}$ gm/cm$^{3}$, where no muons 
exist, the two resulting conditions are:
\begin{equation} \label{eq:nomuons}
   Y_e+Y_{\nu_e}=Y_l=const\;\;\;\; ; \;\;\;\; Y_\mu+Y_{\nu_\mu}=0
\end{equation}
where $Y_l$ is determined at the onset of trapping at about 0.35-0.4, with 
$Y_e/Y_{\nu_e}\approx5-7$, depending on the exact trapping density (see Bethe's 
detailed review \cite{Bethrev}). As metioned earlier, we will ignore the minute muon 
production in proto-neutron-star in this study.	Neutron star matter, on the other hand, 
is completely transparent to neutrinos (the transparency settles after about 50 
seconds of proto-neutron-star evolution \cite{Borr88}), and the neutrino chemical 
potentials are set to zero. Muons are then easier to create (the threshold condition 
for muon creation is simply $\mu_e=m_\mu)$, but the total lepton density is not 
constrained. Neutron star matter is considerably deleptonized with respect to the 
proto-neutron star. The two equilibrium conditions are:
\begin{equation} \label{eq:withmuons}
  \mu_e=\mu_\mu\;\;\;\; ; \;\;\;\; \mu_{\nu_e}=\mu_{\nu_\mu}=0\;\;,
\end{equation}
and all the chemical potentials can be expressed as linear combinations of the 
electron and neutron chemical potentials (associated with the electric and baryonic 
charges, respectively). 

Solving the complete set of conditions yields the equilibrium fractions of the 
baryon and lepton species for any given density and temperature. The effective form 
of our EoS enables a low-order Newton-Raphson iteration process, which is quick and 
efficient. The equilibrium compositions and the resulting thermodynamic properties 
are then calculated for the large variety of strong interaction parameterizations 
discussed below.

Other particle candidates for appearance in the high density matter are the 
$\Delta$ isobars and $\pi$ and K mesons as condensates. Current theories predict 
$\pi$N S-wave repuslive forces which exclude pion condensates, and indirectly, 
make the formation of $\Delta$'s unfavorable as well \cite{Glenn85,Latt2}. However, 
as shown by Weber and Weigel \cite{WebWeig}, in Relativistic-Hartree-Fock calculations 
some evaluations of the $\rho$N$\Delta$ coupling enhance the formation of the 
$\Delta^-$. Kaon and anti-kaon condensation has attracted considerable attention 
\cite{Brnalkaon}, but some recent investigations (see, for example, Refs. 
\cite{SchafMish,Schafalkaon}) do not find kaon effective masses dropping low enough 
at relevant densities. Clearly, the formation of hyperons in the matter will further 
suppress meson condensation \cite{SchafMish,newSB}. We therefore opt not to include 
$\Delta$'s and meson condensates in this study; further generaliztion is, of course, 
possible, where the model dependence of the above arguments may be examined. A final 
issue is that of the transition into a deconfined quark phase: at some high density 
the hadronic identities of the quark states must begin to fade, as the baryons will 
overlap. In this study we do not treat quark degrees of freedom, and 
allow the baryonic effective equation of state to dominate up to 8$\rho_s$. Quark 
degrees of freedom, however, might come into effect at lower values of density (an 
especially exotic case is the possibility of stable strange quark matter, which 
leads to the existence of strange stars \cite{AlFarOlSS}). The issue of quark 
deconfinement is defered to a future work.

\section{Choice of Parameters}
The local potential approach requires four parameters for every combination of 
baryon-baryon interactions in Eq. (\ref{eq:epotbar}): the coefficients $a,b,c$ 
and the density power dependence $\gamma$ ($\delta$ for the nucleon-nucleon case). 
The effective masses of the baryons, or rather their density dependence through the 
parameters $\beta_i$ defined in Eq. (\ref{eq:meffi}), are required as well.

In principle, all parameters of the local potentials could be determined
experimentally, but in practice there are large uncertainties due to limited 
experimental data, and some values must be drawn from theory. The experimental 
and theoretical data used as basis for evaluating the parameters are discussed 
briefly in the proceeding paragraphs.

\subsection{Nucleon-Nucleon Interactions}
The properties of the nucleon interactions at the nuclear saturation density, 
$\rho_s$, are measured to a reasonable precision in finite nuclei. The 
experimental value of this density is estimated to be in the range 0.145-0.17 
fm$^{-3}$, and in this study we follow LS by employing the Skyrme I' force value of 
$\rho_s=0.155$ fm$^{-3}$.	

As shown in LS, the values of the parameters $a_{NN},b_{NN},c_{NN}$ and $\delta$ 
are evaluated from the binding energy, symmetry energy and incompressibility of 
saturated symmetric nuclear matter, derived from typical experimental results 
and the Skyrme model (see Section 2.2 of LS for detail). However, the parameters 
$a_{NN},c_{NN}$ and $\delta$ strongly depend on the nuclear incompressibility 
which is poorly determined in finite nuclei, and is difficult to match with 
theoretical predictions (which generally yield too high incompressibility). 
Thus, quantifying the properties of the nucleon-nucleon interactions at high densities 
leaves a large range of uncertainty, especially for $\delta$, which as a 
result may be set practically arbitrarily between unity and about 2. For example, 
LS use the Skyrme value of the incompressibility $K_s=375$ MeV, leading to 
$\delta=2.002$, but in a later work \cite{SwesLattMy} values down to $K_s=180$
MeV, resulting in $\delta=1.07$ are also considered. In order to investigate the 
model dependence due to this parameter, we will examine the values 
$\delta=2,\frac{5}{3}$ and $\frac{4}{3}$. The appropriate values for the parameters 
$a_{NN},b_{NN}$ and $c_{NN}$ for these values of $\delta$ are presented in Table 1.

\subsection{Hyperon-Nucleon Interactions}
Experimental data of effective nucleon-hyperon interactions are confined to 
hypernuclei (for a recent review, see \cite{ChirDov}). This clearly limits the 
interpretation of the data, and requires a combination of experimental and 
theoretical analysis.

There is a considerable body of information on binding energies, $B_\Lambda$, of 
$\Lambda$'s bound in various single particle orbitals in hypernuclei with total 
baryonic number from $A=3$ to $A=208$. By analyzing these data in a Skyrme-Hartree-Fock 
picture MDG determined a well depth of 
\begin{equation} \label{eq:lonwell}
  -V_\Lambda(\rho_N=\rho_s)\approx27-28 \mbox{ MeV}\;   ,
\end{equation}
defined as the potential seen by a single $\Lambda$ in nuclear matter at saturation 
density, which is roughly $\frac{1}{3}$ of the nucleon well depth in a similar 
parameterization. Constraining the local potential by the well depth at nuclear 
saturation density is the equivalent of using the bulk binding energy to constrain 
the nucleon-nucleon interaction (related to the well depth through the Fermi 
kinetic energy). By combining the data for many hypernuclei, MDG used the $A$ 
dependence to evaluate the $\Lambda$-nuclear potential for densities $\rho\leq\rho_s$ 
and fitted a local potential of the form $V_\Lambda(\rho_N)=a\rho_N+c\rho^\gamma_N$ to 
the experimental results, deducing $a$ and $c$ for a given $\gamma$. MDG achieved 
reasonable fits for the single $\Lambda$ orbitals in ordinary hypernuclei, but - 
again - could not constrain the shape of the potential for the high density regime, 
and therefore could not uniquely determine the exponent $\gamma$ of the repulsion term.
Several possible choices for the parameters $a,c$ and $\gamma$ are presented in Table 
2 below. Figure 1, reproduced from MDG, demonstrates the character of the uncertainties 
of the baryon interactions, as inherent in the local potential formulation. 
The single $\Lambda$ local potential in nuclear matter, $V_\Lambda(\rho_N)$, is plotted 
against the nuclear matter density, for different choices of the exponent $\gamma$. 
All the potentials are constrained by a well depth of about $-28$ MeV at nuclear 
saturation density and the $\Lambda$-hypernuclear binding energies for lower densities, 
so that all the plots yield similar results for $\rho\leq\rho_s$. However, the 
unconstrained behavior at high density is quite divergent, depending strongly on the 
value of $\gamma$. For example, the cross-over density, $\rho_{co}$ for which 
$V_\Lambda(\rho_{co})=0$ varies from about $\rho_{co}\!=\!2\rho_s$ for $\gamma\!=\!2$ 
to $\rho_{co}\!=\!3.5\rho_s$ for $\gamma\!=\!\frac{4}{3}$. It is noteworthy that this 
local potential does reproduce the calculated single particle potentials  found in RMF 
models \cite{Schafpriv}.

Experimental data of $\Sigma$ hypernuclei are more ambiguous, partly because of 
the strong $\Sigma$N$\rightarrow\!\Lambda$N decay channel. It is common to assume 
that the $\Sigma$ well depth is equal to that of the $\Lambda$ \cite{DovGal2}. 
The isospin dependent component of the interaction of a $\Sigma$ hyperon in nuclear 
matter ($\Sigma$-nm) is usualy evaluated from theory, typicaly predicting a $\Sigma$ 
isovector coupling constant (due to the $\rho$ meson field in RMF models) of the 
same order of magnitude as for nucleons. Possible choices for the parameters $a,b,c$
and $\gamma$ then follow from Tables 1 and 2, and are presented in Table 3. However, 
a recent analysis \cite{BattFrGal,MarFrGalJen} of theoretical work and experimental data 
concerning $\Sigma^-$-atoms suggests that there is a strong isoscalar repulsion
in the $\Sigma$-nm interaction. These studies, which also determine the isovector 
component of the $\Sigma^-$-nucleus interaction, indicate that $\Sigma$ baryons 
are strongly repulsed in the nuclear interior. A corresponding density dependent local 
potential has also been proposed (again with difficulty in determining the high power 
density dependence) for $\Sigma^-$ hyperons:
\begin{equation} \label{eq:signopt}
  2\mu V_\Sigma(\rho_N)=
        -4\pi\left(1+\frac{\mu}{m}\right)
        \left\{\left[b_0+B_0\left(\frac{\rho_N}{\rho_s}\right)^\alpha\right]
        +\left[b_1+B_1\left(\frac{\rho_N}{\rho_s}\right)^\alpha\right]
        (n-z)\right\}\rho_N\;  ,
\end{equation}

where $\mu$ is the $\Sigma$-nucleus reduced mass (for nuclear matter it is simply 
the $\Sigma$ mass) and $m$ is the nucleon mass. The $(b_0,B_0)$ term describes the 
isoscalar component of the interaction, and with $B_0\!<\!0$ and $|B_0|\!>\!b_0$ a 
strong repulsion is produced in the nuclear interior. The second term introduces 
isospin dependence, reversing sign for the $\Sigma^+$. The real part of this potential 
may be taken as the $\Sigma$-nm interaction for equilibrium purposes, as the imaginary 
part represents mainly the $\Sigma$N$\rightarrow\!\Lambda$N channel. A representative 
set of parameters for Eq. (\ref{eq:signopt}) is given in Table 4. We investigate both 
of the possibilities listed in Tables 3 and 4 for the $\Sigma$-nm interaction in 
constructing the equation of state.

In emulsion experiments with K$^-$ beams there are a few events attributed to the 
formation of $\Xi$-hypernuclei. Although the evidence is far from being compelling, 
the data can be interpreted \cite{DovGal1} consistently in terms of a potential well 
depth of
\begin{equation} \label{eq:hypwell}
  -V_\Xi(\rho_N=\rho_s)\approx20-25\mbox{ MeV}\;   .
\end{equation}
Such a result has also been derived in one-boson-exchange (OBE) models \cite{DovGal2}. 
The isospin component of the $\Xi$N interaction is very model dependent, ranging from 
negligible \cite{DovGal2} (an order of magnitude smaller than that observed for 
nucleons) to equal to that of nucleons \cite{SchafMish}. Possible choices for the 
parameters $a,c$ and $\gamma$, assuming $b_{\Xi N}\!=\!0$ are listed in Table 5. In 
deriving these parameters we assumed $-V_\Xi(\rho_N\!=\!\rho_s)=24$ MeV for symmetric
nuclear matter (where symmetry terms vanish) and that for a given value of $\gamma$ the 
crossover density, $\rho_{co}$, for which $V_\Xi(\rho_N\!=\!\rho_{co})=0$ is equal to 
that for $\Lambda$ in nuclear matter, thus simulating universality in the high density regime.

\subsection{Hyperon-Hyperon Interactions}
The experimental data regarding hyperon-hyperon interactions is extremely scarce. 
There are a few events which may be interpreted by the creation of $\Lambda\Lambda$ 
hypernuclei (see Ref. \cite{ChirDov}). Analysis of these events requires a 
rather strong $\Lambda\Lambda$ attraction, which is not easily reproduced by theory 
\cite{BodUsm}. Recent studies with RMF models remedied the hyperon-hyperon interaction 
by adding two mesons \cite{SchafMish,SchafGalalSHM} which couple only to the hyperons. 
Schaffner et al. \cite{SchafGalalSHM} constructed single particle potentials on the 
basis of OBE calculations of the Nijmegen group, and estimated the well depths of 
hyperons in hyperon matter as follows:
\begin{equation} \label{eq:YYwell}
  -V_{h_1}(\rho_{h_2}\!=\!\rho_s)=40\mbox{ MeV}\;\;   ,
\end{equation}

implying a universal hyperon-hyperon interaction. Similar to the case of the $\Xi$-nm
interaction, the coefficients $a_{h_1h_2}$ and $c_{h_1h_2}$ can be determined for a 
given value of $\gamma$ by requiring that for symmetric matter the 
potential depth at a density equal to the nuclear saturation density be equal to the 
value assumed in Eq. (\ref{eq:YYwell}), and that the crossover density for which the 
potential is zero, be equal to that of the $\Lambda$ in nuclear matter ($\Lambda$-nm) 
case. The appropriate values for the inter-hyperon interactions are given in Table 6.

Generally, the symmetry coefficients $b_{xy}$ in Eq. (\ref{eq:epotbar}) are independent 
of $\gamma$, and must be determined separately. The isospin dependent components of the 
hyperon-hyperon interaction must be drawn from theory. As mentioned earlier, the 
isovector coupling constant of the $\Sigma$ baryons is typically assumed to be up to 
twice that of the nucleons, while that of the $\Xi$ baryons is unclear.
We assumed that only the N-nm, $\Sigma$-nm and $\Sigma$-$\Sigma$m interactions have a
nonzero isospin component, and the corrseponding values of the symmetry coefficients are
presented in Tables 1, 3 and 6. 

\setlength{\parskip}{0.3in}

We conclude this discussion by reemphasizing the large uncertainties surrounding 
the baryon-baryon interactions, especially regarding the density dependence. 
Constraining the local potentials in nuclear matter by assuming a given well 
depth at saturation density does not limit the shape of the potential at 
higher densities, which is handled in the local potential forms of Eq. 
(\ref{eq:epotbar}) by the parameters $\gamma$ and $\delta$. We therefore treat the 
exponents $\gamma$ and $\delta$ as free variables, and simply assume a universal 
nature to all multibody forces, so that we may also set $\gamma\!=\!\delta$.

The evaluation of the baryon effective masses is a fundamental problem in any analysis
of supernuclear density matter. The values of effective masses are needed for the 
kinetic and mass energy density terms of Eqs. (\ref{eq:ekin}) and (\ref{eq:emass})
respectively, and nuclear matter calculations typically predict \cite{Vauth} that the 
nucleon mass in bulk matter is smaller than the bare mass by a factor of 
$m^*/m\approx0.7$ at saturation density, and will continue to 
decrease for higher densities, due to both interaction and relativistic effects. 
The analysis of the effective masses of all the baryons in \cite{SchafMish} yields 
similar results for nucleons and hyperons alike.

\setlength{\parskip}{0.1in}

The local potential has no mechanism to generate both the interactions and the 
effective masses consistently. LS chose to set  $m^*/m=1$, which is a reasonable 
approximation for densities up to about twice the nuclear saturation density 
(and may also be justified by some uncertainties regarding effective masses at the 
Fermi surface \cite{Expdat}). At higher densities such an approximation becomes 
questionable. While the explicitly density dependent interactions in the potential 
model may compensate for some of the expected decrease in the effective masses (LS 
include a further explicit dependence of the forces on the assumed values for the 
effective masses - see Eqs. (2.20-21) in LS), relativistic effects are not taken 
into account. 

For lack of a better model, in this work we will use $m^*/m=1$. An obvious flaw in 
setting all effective masses equal to the bare ones is that mass differences between 
the various types of baryons remain constant. Theory clearly suggests otherwise 
(see, for example, Figs. 4 and 7 in \cite{SchafMish}), and changing the mass 
differences will clearly affect the equilibrium compositions. For example, theory 
predicts that the effective mass of nucleons in nuclear matter at $\rho\!=\!\rho_s$ is 
$m^*_N/m_N\!\approx\!0.55-0.7$, while that of the $\Lambda$ decreases only to 
$m^*_\Lambda/m_\Lambda\!\approx\!0.8$ \cite{MDG}; hence the mass difference will be 
larger than for the bare masses (and $\Lambda$'s will form in nuclear matter at higher 
densities than for $m^*/m\!=\!1$). The sensitivity of the results to this crude 
assumption can be examined by using nonzero positive values for the coeffecients 
$\beta_i$ of Eq. (\ref{eq:meffi}), which will yield effective mass functions 
similar to those found in RMF models (for example, in Refs. \cite{SchafMish,Latt1}).

\section{Results: Equilibrium Compositions and Equations of State}

The basic features of the dense matter are the different particle fractions and 
the equation of state, both of which are dependent on the baryon density of the 
matter, and of the modeling of the strong intearctions. In this Section the 
equilibrium compositions and equations of state are examined and compared for the 
different cases of candidate hyperon species (no hyperons, $\Lambda$'s, $\Sigma$'s 
and $\Xi$'s, and only $\Lambda$'s and $\Xi$'s but no $\Sigma$'s when assuming a 
repulsive $\Sigma$-nm interaction). The sensitivity to the choice of parameters 
of the baryon-baryon interactions is also examined, by setting different values for 
the exponents $\delta$ and $\gamma$.

We shall focus on the simpler case of neutron star matter, which can be well 
approximated as being at zero temperature, where all thermodynamic functions are 
easily found. Some results regarding proto-neutron-star (core-collapse) matter are 
presented in the latter part of this Section, where the assumed thermodynamic 
conditions are detailed.

\subsection{Neutron Star Matter}

Neutron star matter may be well approximated as having zero temperature, since 
evolved neutron stars have internal temperatures considerably below 1 MeV. The internal 
energy density  $\varepsilon$ and the free energy density $f$ are identically the 
same, and the kinetic energy terms in Eqs. (\ref{eq:ekin},\ref{eq:taui}) are 
all simply related to the Fermi momenta.

Shown in Fig. 2 are the particle fractions versus the baryon density for matter 
containing neutrons, protons electrons and muons, calculated by using the parameters 
for $\delta\!=\!2$ (see Table 1), and ignoring the possible appearance of hyperons. 
Compositions are calculated up to about $8\rho_s$, where quark degrees of freedom are 
assumed to come into effect, and the validity of a baryonic model of the matter becomes 
questionable. These plots are very similar to those calculated in various models of 
nuclear matter for neutron stars. The main features of such matter are:
\setlength{\parskip}{0.0in}
\begin{enumerate}

\item Neutrons are by far the dominant species in the matter.
\setlength{\parskip}{0.0in}

\item Both electrons and muons are present in the matter. At increasing 
densities the chemical potential of the leptons rises (due to the rising Fermi 
energies), and the muon fraction approaches  that of electrons, as the Fermi 
energies drive the chemical potential up so that the mass difference is of lesser 
significance.
\setlength{\parskip}{0.0in}

\item The balance of nucleons and leptons is determined by the difference 
$\mu_n-\mu_p$ (commonly denoted as $\hat{\mu}$) and by $\mu_e$. Equilibrium 
demands that both be equal, but since $\mu_n-\mu_p$ has a steeper density 
dependence, there is some decrease in the neutron fraction and some increase in the 
proton and total lepton fractions as a function of the total baryon densities.
\end{enumerate}
\setlength{\parskip}{0.1in}

These features are common to a variety of nuclear matter models and are only weakly 
model dependent. This results in similar graphs for the other parameterizations; 
although specific details are, of course, model dependent (for example, the 
exact proton fraction which is highly important for direct Urca cooling rates).

The picture clearly changes when hyperons are taken into account. Figures 3a, 3b and 
3c show the various particle fractions as a function of the total baryon density for 
$\gamma\!=\!\delta\!=\!2,\frac{5}{3}$ and $\frac{4}{3}$ respectively. 
In all cases the hyperon-hyperon well depth was set as $-40$ MeV, and the $\Sigma$-nm 
interaction is assumed to be equal to the $\Lambda$-nm one (except for the symmetry 
term, which is nonzero for $\Sigma$'s and nucleons). The appearance of hyperons 
affects all the main features of the equilibrium compositions. The key effects to 
be noted are:
\setlength{\parskip}{0.0in}
\begin{enumerate}
\setlength{\parskip}{0.0in}

\item In all cases hyperons appear at about twice the nuclear saturation density, 
and at densities of about $2.5\rho_s$ they sustain a significant fraction of the 
total baryon population. 
\setlength{\parskip}{0.0in}

\item The first hyperon species to appear is the $\Sigma^-$, closely followed by 
the $\Lambda$. The negative charge of the  $\Sigma^-$ outweighs the 80 MeV mass 
difference, as a result of the more lenient condition, Eq. (\ref{eq:bar_eqlbm}),
for its chemical potential. However, the formation of $\Sigma^-$ hyperons is 
quickly moderated by the charge dependent forces, which disfavor an excess of 
$\Sigma^-$'s over $\Sigma^+$'s, and a joint excess of $\Sigma^-$'s and neutrons. 
Thus the $\Sigma^-$ fraction saturates at about 0.1, while the $\Lambda$'s, free of 
charge-dependent forces, continue to accumulate until the repulsive (multibody) 
forces bring saturation on them as well.
\setlength{\parskip}{0.0in}

\item All other hyperon species follow, and appear one by one in the matter. With 
the assumptions mentioned above, $\Sigma$'s generally appear before the $\Xi$'s due 
to the large mass difference, but the $\Xi^-$ becomes favored due to its negative 
electric charge, and quickly becomes abundant in the matter. Since we assume no 
charge-dependent forces for the $\Xi$ baryons, the $\Xi^-$ fraction does not 
saturate quickly, and its fraction overtakes that of the $\Sigma^-$ at high 
densities.
\setlength{\parskip}{0.0in}

\item A unique aspect of hyperon appearance in all cases is the immediate 
deleptonization of the matter. Leptons are rather expensive in terms of energy 
density (and pressure), surviving in nuclear matter only because of the need for 
charge neutrality and because the charge-dependent forces impose a significant 
proton fraction. Hyperons create an option for lowering the neutron excess free of 
lepton formation, and the negatively charged hyperons allow charge neutrality to be 
maintained within the baryon community. Thus, the increase of the lepton abundance 
is cut short by the formation of the hyperons and, in some cases, the appearance of 
the $\Xi^-$ is followed by a very powerful deleptonization. For $\gamma\!=\!\delta\geq\frac{5}{3}$ the muon population is even completely extinguished (obviously, the 
$\Sigma^-$s cannot achieve this alone because of their saturation fraction, imposed 
by the symmetry forces). The case of $\gamma=\delta=\frac{4}{3}$ is somewhat 
eccentric, maintaining a high lepton content along with the rather moderated 
appearance of hyperons. This is a consequence of setting $\delta\!=\!\frac{4}{3}$, 
which represents a too mild increase in the nucleon-nucleon repulsion with increasing 
density, thus suppressing hyperon production. Such a composition seems to be 
unlikely, since a combination of hyperons and a high lepton content will create very 
rapid cooling rates [32-35].	
\end{enumerate}
\setlength{\parskip}{0.1in}

Some model dependence does exist, of course, as weaker nucleon repulsion is 
necessarily accompanied by less hyperon production. For $\delta=2$, hyperon 
production is so favorable that at high densities $\Lambda$'s, not neutrons, are the 
most abundant species, while for $\delta=\frac{4}{3}$ nucleons dominate the matter 
even at $8\rho_s$. The appearance of hyperons other than $\Sigma^-$ and $\Lambda$ 
is also somewhat model dependent: the more powerful the high density 
repulsion (among nucleons and hyperons), the formation of $\Sigma^0$  and even 
$\Sigma^+$ becomes more favorable at sufficiently low densities, where the more 
massive $\Xi^-$ is still absent; for weaker repulsion, it is the $\Xi^-$ which 
appears first of the three, at densities where the effect of its negative charge is 
sufficient to overcome the 120 MeV mass difference. $\Xi^0$'s appear only when 
repulsive forces become dominant again and the introduction of more species is thus 
more effective in lowering the potential energy density.

The reason for the hyperon formation density being only weakly model dependent is 
that when hyperons are excluded, the nucleon chemical potentials in nuclear matter
rise steeply with density. All models predict that at about $2\rho_s$ the nucleon-nucleon 
interaction is far beyond maximum attraction, and in some models it is already 
repulsive, so that lowering the partial density of the nucleons is clearly favorable. 
Thus, although the exact density for hyperon appearance is model dependent, it does 
not vary by more than a few hundredths of a fm$^{-3}$. We note that this is also 
the reason for the quick rise in the hyperon fractions with density after formation. 
The only balance to this trend are the kinetic energies, which are far outweighed until 
the hyperon fraction rises to the order of about 0.1. This behavior is less 
pronounced for the $\Sigma$'s than for the $\Lambda$'s (and in our model, also for 
the $\Xi$'s) because of the nonzero symmetry term, which disfavors a fast build-up 
of one species.

When using the effective equation of state, the effects discussed above tend to 
overshoot, creating non-monotonic irregularities in the curves (most notable in the 
proton and lepton fractions). This (unphysical) effect is moderated to large extent 
in RMF models, which add further control through the effective masses which are 
density and species-fraction dependent.

Generally, Figs. 3a and 3b show a fair agreement with the results of the 
equilibrium compositions found in the mean field calculations of [3-5]. This agreement 
suggests that the effective equation of state can indeed be considered as a trustworthy 
tool in the analysis of high density baryonic matter. The somewhat eccentric case of 
$\gamma=\delta=\frac{4}{3}$ examined in Fig. 3c, however, is qualitatively different 
from the RMF results. It seems that this case should be treated with some 
suspicion, although we will keep it in the analysis below. It is noteworthy that the 
quantitative details of the compositions are model dependent in the RMF calculations 
as well. For example, the deleptonization rate and other features of the PL-Z model in 
Ref. \cite{SchafMish} resemble our $\gamma\!=\!\delta=2$ case, while those of case 2 in 
Ref. \cite{Glenn85} (which is relevant for our discussion) bear higher resemblance to 
the  $\gamma\!=\!\delta\!=\!\frac{5}{3}$ case.

We now turn to the case of strong isoscalar repulsion in the $\Sigma$-nm interaction, 
suggested in \cite{BattFrGal,MarFrGalJen}. Using the local potential of Eq. 
(\ref{eq:signopt}) with the parameters listed in Table 4 (or any of the other sets of 
parameters in \cite{BattFrGal} that fit the experimental data of $\Sigma^-$-atoms) 
yields a repulsive potential of the order of tens of MeV's for $\Sigma$'s in nuclear 
matter at saturation density, and even greater repulsion at higher densities. $\Sigma$ 
hyperons are thus disfavored in nuclear matter, and the obvious byproduct is the 
exclusion of these species in the high density matter under discussion.

Figures 4a, 4b and 4c show the various particle fractions as a function of the total 
baryon density for $\gamma\!=\!\delta\!=\!2,\frac{5}{3}$ and $\frac{4}{3}$ 
respectively, when the hyperon-hyperon potential well depth is set as $-40$ MeV, and the 
$\Sigma$-nm interaction is constarined by fitting to the $\Sigma^-$-atoms data. Clearly, 
$\Sigma$ hyperons are absent, but other than this, the key features of hyperon 
formation are unchanged:
\begin{enumerate}
\setlength{\parskip}{0.0in}

\item Hyperons appear again at about twice the saturation density, and at about 
$2.5\rho_s$ the hyperon fraction is already significant. In this case, however, the 
entire effect is due to formation of $\Lambda$'s. This quantitative result is 
practically model independent. The appearance of the $\Lambda$'s halts the increase 
in the lepton population, and when $\Xi^-$'s are added further deleptoniztion occurs. 
The $\Lambda$ actually appears at lower densities than it did when $\Sigma$'s were 
present, since now it has no competition from the $\Sigma^-$.

\setlength{\parskip}{0.0in}
\item The $\Xi^-$ appears as well in all cases, at slightly lower densities than for the corresponding case when competing with $\Sigma$'s. The formation of $\Xi^0$ is 
model dependent, again, since stronger high density repulsion favors the introduction 
of more species.
\end{enumerate}
\setlength{\parskip}{0.1in}

The total hyperon fraction is again highly model dependent as are the specific 
details of the deleptonization.

The sensitivity of the results to the assumed value of the hyperon-hyperon well 
depth, is examined by setting its value to zero - ``turning off'' the hyperon-hyperon 
interaction (with the exception of the $\Sigma$-$\Sigma$m symmetry term, which is 
kept unchanged). Fig. 5 presents the equilibrium compositions found for the 
$\gamma\!=\!\delta\!=\!2$ case when all baryon species are allowed (assuming the 
$\Sigma$-nm interaction to be equal to the $\Lambda$-nm one) for a zero hyperon-hyperon 
well depth (while all other parameters are unchanged). The considerable similarity to 
the compositions for the corresponding case with a relatively large hyperon-hyperon 
well depth (Fig. 3a) is clear. There is a slight moderation in the rate of 
accumulation of hyperons (missing the attractive interactions among hyperons of the 
same species at small partial densities) while the appearance of the more massive 
hyperons is somewhat delayed, because there is also no repulsion among the abundant 
hyperons. These effects are visible but clearly are of second order, since the matter 
is dominated by nucleons in most of the density range, so it is the nucleon-nucleon 
and hyperon-nucleon interactions which determine the equilibrium compositions. The 
sensitivity to the hyperon-hyperon interaction is even weaker for lower values of 
$\delta$ and $\gamma$, as the nucleon fraction in these cases is larger. A similar 
relative insensitivity to the hyperon-hyperon well depth is found also when 
$\Sigma$'s are excluded from the matter.

The solution for the equilibrium compositions enables the 
determination of the thermodynamic functions, explicitly the energy density and the 
pressure, for constructing the EoS. The key point regarding the effect of 
introducing extra particle species into the matter in this respect is the 
softening of the EoS. This softening is the consequence of turning potential and 
kinetic energy of the already present baryons into masses of the new species, and in 
reducing repulsion (and sometimes even gaining attraction) in the potential energy 
terms, due to the lowering of the partial densities of the abundant species. These 
two effects lower the energy density (so that the mass-energy density at a given 
baryon density is somewhat less than the corresponding energy density of nuclear 
matter at the same baryon density), but the main effect involves the pressure, since 
storing energy in mass is cheapest in terms of pressure. The decrease of the 
lepton population does add further softening, but this effect is of second order.

The nature of the EoS is thus dependent on the number of species, in addition to the 
obvious dependence on the baryon-baryon interaction model. It is instructive to begin 
with the resulting EoS for nuclear matter for the three cases corresponding to 
$\delta\!=\!2,\frac{5}{3}$ and $\frac{4}{3}$, plotted in Fig. 6. Since no hyperons 
are considered, the resulting EoS is a direct function of the nuclear 
interaction model, namely, of the value of $\delta$. The differences between the three 
equations are clearly visible. It is noteworthy that the $\delta\!=\!2,\frac{5}{3}$ 
equations break the causality limit $(\left(dP/d\varepsilon\right)\leq c^2)$ at some 
density, where the plots are made dashed. This is a known flaw in effective models for 
the nuclear interactions, sometimes amended by introducing a finite-range into the 
local potentials.	

The properties of the EoS change qualitatively when hyperons are taken into account: 
Fig. 7 shows the resulting EoS for three types of matter with 
$\delta\!=\!\gamma\!=\!\frac{5}{3}$, for a hyperon well depth of 
$-V_{h_1h_2}\!=\!40$ MeV. The three cases are 1) nuclear matter (n-p-e-$\mu$); 2) 
baryon matter including $\Lambda$'s and $\Xi$'s but no $\Sigma$'s (assuming a repulsive 
$\Sigma$-nm interaction \cite{BattFrGal}); and 3) all eight 
species of baryons appear. The three plots demonstrate clearly the softening of the 
EoS when hyperons are included, this softening becoming more pronounced 
when the $\Sigma$'s are also included.

We must note that the numerical peaks in the particle fractions discussed above are 
unfortunately carried over to the EoS, when the unphysical fast rise in 
the $\Lambda$ fraction (and sometimes the $\Xi^-$ fraction as well) yields a slight 
dip in the pressure, so that over a small region the unphysical result of $\partial 
P/\partial\varepsilon<0$ is obtained. By employing a mathematical smoothing scheme, 
interpolating the pressure-density dependence in these small regions, we are able to 
amend this shortcoming, particullary for the case of $\gamma\!=\!2$.

Fig. 8 compares the resulting equation of state for three cases of the 
baryon-baryon interactions, i.e. $\gamma\!=\!\delta\!=\!2,\frac{5}{3}$ and 
$\frac{4}{3}$, and in all cases the hyperon-hyperon well depth is assumed to be 
$-V_{h_1h_2}\!=\!40$ MeV, and the $\Sigma$-nm interaction is assumed to be 
equal to the $\Lambda$-nm one. It is noteworthy that all three equations of state 
are quantitatively similar, indicating a rather weak dependence on the precise 
description of the short-range forces represented by the exponents $\delta$ and 
$\gamma$. A similar trend is found when a highly repulsive $\Sigma$-nm interaction is 
assumed, with all equations being somewhat stiffer than the coresponding cases where
the $\Sigma$'s are allowed in the matter. Thus, an important feature of our results 
is that the equation of state is much less sensitive to the details of the 
baryon-baryon interaction, with the obvious exception of the two extreme 
possibilities for the $\Sigma$-nm interaction. 

The nucleon-nucleon interaction, which determines the stiffness of the nuclear-matter 
equation of state, also determines the rate at which the nucleon chemical potential 
rises  with increasing density. A greater rise in the nucleon chemical potential is 
the incentive to enhanced formation of hyperons in the matter, which in turn softens 
the total equation of state. {\em Our key observation is that the macroscopic 
properties of the equation of state are thus strongly coupled to the microscopic 
equilibrium compositions, and once hyperons are available the resulting equation of 
state is constrained to a significantly narrower range of values}. We may conclude 
that {\em hyperons serve as a ``pressure control'' mechanism} in high density matter, 
and its nature depends mainly on the number of hyperon species available.

Explicitly, as long as attractive nucleon-nucleon and nucleon-hyperon well depths are 
assumed, the resulting EoS is only mildly sensitive to the choice 
of $\delta$ and $\gamma$. This is mainly reflected in the width of the 
"semi-plateau" region, where the pressure grows only very mildly: the greater the 
value of the exponents, the higher the relative gain $|dV/d\rho|$ in attraction at 
low densities and the wider the density range where the hyperon appearance, mainly of 
the $\Lambda$, creates the "semi-plateau" region. On the other hand, if a large value 
is chosen for the exponents, there is also a steeper rise of repulsion 
at higher densities, which builds up the pressure faster. These two effects are 
balanced by the appearance of new species in the matter, when such species are 
available. New species will appear when the pressure (and internal energy) rise 
steeply with the baryon density, but their appearance is delayed when the species 
already present moderate the pressure build up, as happens in the "semi-plateau" 
regions.

The softening of the EoS may be demonstrated by calculating the 
adiabatic index $\Gamma=\varepsilon/P\partial(P)/\partial(\varepsilon)$. Fig. 9 shows the 
adiabatic index as a function of the mass energy density, for nuclear matter and for
matter containing nucleons and hyperons, with $\delta\!=\!\gamma\!=\!\frac{5}{3}$. 
The appearance of the hyperons is easily identified, as each of them 
lowered the adiabatic index as long as its fraction did not saturate. The 
$\Lambda$'s and $\Xi$'s cause greater ``dips'' in the value of $\Gamma$ than the 
$\Sigma$'s, since their formation is not moderated by charge dependent forces.

The sensitivity of the ``pressure control'' to the hyperon-hyperon interaction is 
found to be especially weak, since the matter is nucleon dominated in at least 
most of the relevant density range (and for $\gamma\!=\!\delta\!=\!\frac{4}{3}$ even 
for the entire range). The deeper $V_{h_1h_2}$ is, there will 
be a stronger hyperon-hyperon attraction at low partial densities (further favoring 
hyperon production), but also a stronger hyperon-hyperon repulsion at high partial 
densities. Hence, through most of the density range the hyperon fractions quickly 
saturate so that their contribution to the energy density (and pressure) are 
relatively independent of the details of the hyperon-hyperon interaction.

We comment briefly on the issue of the effective masses. Several equilibrium 
compositions were recalculated with baryon effective masses $m^*/m\leq1$ derived by 
using nonzero values for the coeffecients $\beta_i$ in Eq. (\ref{eq:meffi}). We 
find that as long as the mass differences between the various baryon species do not 
change considerably from the case of bare masses, the equilibrium compositions are very 
similar to those found by using $m^*/m\!=\!1$. This is a natural result, since the 
chemical potentials are highly density dependent through the potential energy terms, 
and so the densities at which hyperons appear do not vary by more than a few hundredths 
of fm$^{-3}$. The corresponding equations of state are more sensitive to the values of 
the effective masses, requiring some careful consideration. We find that the pressure 
vs. baryon density is only mildly dependent on the precise choices for $m^*/m$, and the 
basic feature of hyperon induced ``pressure control'' is maintained. Since the mass 
density gives a large contribution to the energy density, the pressure vs. energy density 
relation is more complicated when effective masses are used (and, indeed, LS scale 
their energy density relative to the neutron effective mass). However, if the energy 
density is scaled so that for nuclear matter at nuclear saturation density it yields 
the typical value of $\varepsilon(\rho_N\!=\!\rho_s)\approx 2.4\times10^{14}$ gm/cm${^3}$, 
the resulting equations of state are once again approximately similar to those found 
for $m^*/m\!=\!1$. In summary, we believe that the equations of state found using 
$m^*/m\!=\!1$ reasonably reflect the gross properties of the equations of state that 
are calculated with finer approximations regarding the effective masses.

\subsection{Proto-Neutron Star Matter}

The physical conditions in proto-neutron star matter basically differ from those in 
fully-evolved neutron star matter in two main features: the temperature is of the 
order of 10 MeV and more, so it must be treated as finite, and - to a reasonable 
approximation - the matter is opaque to neutrinos, trapping them and keeping the 
total lepton fraction, $Y_l$, constant.

Both of these general considerations affect the equilibrium composition as well as 
the equation of state of the high density matter. Finite temperature means also 
finite entropy, which will stiffen the EoS of matter for a given composition 
\cite{Latt1} and broaden the population levels of the baryon species, thus lowering 
the chemical potentials. A constant lepton fraction imposes a much higher electron 
fraction than in neutron star matter, clearly affecting the baryon species 
composition, and through it the EoS. Generally, studies of core-collapse and 
proto-neutron matter with finite temperature and a constant lepton fraction point to
the latter as the more important of the two. Its value is determined by the onset of 
neutrino trapping in the collapsing core of the supernova, and is usually found to be 
$Y_l=0.3-0.4$ \cite{Bethrev}; in this Section we use the value of 0.36. The 
temperature is typically constrained by the entropy per baryon, $\tilde{s}$, which 
tends to be constant throughout the star. We use a temperature of $T=20$ MeV, which is 
about equivalent to $\tilde{s}\approx2$ \cite{Latt1}.

Once more we begin with nuclear matter, employing the different assumptions on the 
nuclear EoS. Fig. 10a shows the equilibrium composition of n-p-e-$\nu_e$, with the 
exponent $\delta$ of the nuclear EoS set at $\delta\!=\!\frac{5}{3}$. The basic 
feature of the equilibrium composition is a large electron fraction, imposed by the 
condition that $Y_e+Y_{\nu_e}=Y_l$. The high electron fraction imposes an equal proton 
fraction, so proto-neutron star matter is much more symmetric at $\rho\approx\rho_s\; 
(Z/A\approx0.3)$ than neutron star matter $(Z/A\approx0.1)$. It is also noteworthy 
that the neutrino fraction decreases with density (and, correspondingly, the electron 
fraction increases). This is the result of the equilibrium condition 
(\ref{eq:nu_eqlbm}), and since $\mu_n-\mu_p$ rises more rapidly than 
$\mu_e-\mu_{\nu_e}$, the equilibrium is maintained by lowering the neutron fraction. 
These trends are highly model independent, and similar equilibrium compositions are 
found for the other values of the exponent $\delta$.

Equilibrium compositions of finite temperature matter will always contain some 
finite population of all particle species, so minute hyperon fractions exist even at 
$\rho_s$. However, these hyperon fractions will build up to significant values only 
when the potential and kinetic energies of the nucleons make the transition of nucleons 
to hyperons energetically favorable. Hence, the general features of proto-neutron star 
matter will be qualitatively similar to those of neutron-star matter discussed above, 
and significant hyperon formation is expected.

An example for the effects of neutrino trapping and finite temperature on the 
equilibrium compositions is plotted in Fig. 10b, for the case $\gamma\!=\!\delta
\!=\frac{5}{3}$ and $-V_{h_1h_2}\!=\!40$ MeV, for $Y_l\!=\!0.36$ and a temperature of 
$T\!=\!20$ MeV. These compositions appear to be generally similar to those of the 
corresponding case for neutron star matter (Fig. 3b). Some noticeable differences 
do still exist, of course: the greater symmetry of the nuclear matter at low 
densities, imposed by the large electron fraction, reduces the neutron chemical 
potential with respect to neutron star matter at the same densities. As a result, 
significant hyperon formation is somewhat delayed in the neutrino trapped 
matter. The finite temperature slightly enhances this effect, since the nucleon 
chemical potentials drop below the corresponding value in cold matter. Details
relevant to the hyperon formation are as follows:

\begin{enumerate}
\setlength{\parskip}{0.0in}

\item The formation of the negatively charged hyperons is substantially
suppressed with respect to that in neutron-star matter, since charge neutrality 
is handled by the electrons, which must maintain a sizable fraction throughout 
the entire density range.
\setlength{\parskip}{0.0in}

\item The formation of the $\Sigma^-$ gets delayed, and thus the $\Lambda$ is the 
most favorable hyperon to be produced at the lower densities. The lack of 
competition from the $\Sigma^-$ helps the $\Lambda$, precisely as for the neutron 
star matter case when the $\Sigma$'s were excluded by assuming repulsive $\Sigma$-nm 
interactions. As a result, we find that $\Lambda$'s appear first in the matter, at 
slightly lower densities than when produced in neutron star matter, and it is the 
$\Sigma^-$ which comes in a close second. A similar effect governs the fractions of 
the $\Sigma^0,\Sigma^+$ and $\Xi^0$ with respect to the $\Xi^-$.

\item The total hyperon fraction continues to be smaller than in neutron star 
matter through the enitre density range, although the $\Lambda$'s still accumulate 
to a considerable fraction.

\item The effect of hyperon formation on the lepton population is also worth 
attention. Once again, the formation of hyperons makes possible the reduction of the 
neutron fraction without a coupled growth in the proton fraction. Thus the difference 
$\mu_n-\mu_p$ decreases, forcing a similar trend in $\mu_e-\mu_{\nu_e}$. The 
immediate result is that hyperon formation is closely followed by a reverse of the 
trend in the electron and neutrino populations, with respect to those of nuclear 
matter. The neutrino population begins to rise, and quickly overtakes its value at 
nuclear saturation density of about 0.06 per baryon, and even crosses 0.1 per baryon 
at high densities.
\end{enumerate}
\setlength{\parskip}{0.1in}

Again, the results found here with the effective EoS bear high resemblance to those 
reported in works which used more sophisticated models \cite{Latt1}.

The above results depend only little on temperature, and similar equilibrium 
compositions are found  for any significant temperature $(T\geq10$ MeV). The 
temperature determines the precise hyperon populations at low densities, but these are 
insignificantly low anyway. There is some model dependence with respect to the 
parameterization of the baryon-baryon effective interaction (the values of $\gamma$ 
and $\delta$), in a similar fashion as for neutron star matter. Again, very little 
dependence is found regarding the value of the hyperon-hyperon well depth, which is 
of even less importance in the present case, as the matter is even more nucleon 
dominated.

Similar results are found for the case of a repulsive $\Sigma$-nm interaction, 
excluding $\Sigma$'s from the matter. The equilibrium compositions for proto-neutron-star 
matter with $\gamma\!=\!\delta\!=\!\frac{5}{3}$ and $-V_{h_1h_2}\!=\!40$ MeV, 
but when the $\Sigma$'s are excluded, are plotted in Fig. 10c. Since the formation 
of the $\Sigma^-$ is suppressed even when allowed, $\Lambda$'s dominate inherently and 
the $\Lambda$ fraction is rather insensitive to the presence of $\Sigma$'s. Excluding 
them does create an increase, obviously, in the $\Xi$ fraction.

The outstanding point of proto-neutron star composition is that it generates a 
different equation of state than for neutron star (neutrino-free) matter. In nuclear 
matter the effect is trivial: the greater symmetry among the nucleons lowers the 
total Fermi energies and the symmetry potential energy density. 
Hence, equilibrium nuclear matter with trapped neutrinos is inherently softer than 
neutrino-free nuclear matter for the same nuclear-force model.

The picture is quite the opposite when hyperons enter the game. As discussed above, 
the more symmetric nuclear fraction somewhat suppresses hyperon formation in 
neutrino-trapped matter. Since hyperon formation imposes a softer equation of state by 
lowering kinetic and potential energies, it is clear that when hyperons are allowed to 
appear the equation of state for neutrino-trapped matter, which has a smaller hyperon 
fraction will be stiffer than the EoS for neutrino free matter. This principle was pointed 
out in [7,8], and is laid out in detail in \cite{Latt1}. The equations of state 
calculated here do indeed reproduce these expected results. Fig. 11 compares pressure 
vs. baryon density (which is the relevant quantity when using the same model for 
different thermodynamic conditions), for $\delta\!=\!\gamma\!=\!\frac{5}{3}$ and  
$-V_{h_1h_2}\!=\! 40$ MeV. As is expected for nuclear matter (Fig. 11a) the neutrino-trapped 
case (dashed line) is clearly softer (except at very low densities, where the 
finite-temperature lepton pressure is of significance).  Matter with hyperons presents the 
opposite state of affairs (Figs. 11b and 11c): as is clearly seen, the neutrino-trapped 
cases do indeed have stiffer equations of state than the neutrino-free ones (solid lines). 

\section{Results: Respective Neutron Star Masses}

A natural test for the constructed equations of state is the calculation of the 
neutron star properties they predict. Most important of these is, of course, the 
mass/centeral-density relation: the mass sequence $M(\varepsilon_c)$, of the neutron star 
for a given equation of state, and especially the maximum mass predicted by the 
equation. This maximum mass must be at least $1.4\:M_\odot$, which is the well 
determined observed mass of the Hulse-Taylor pulsar (see \cite{Cook} for a recent 
review of observed and calculated neutron star properties), where $M_\odot\!\equiv\!
1.989\times10^{33}$ gm is the standard notation for one solar mass.

In this study we limit our discussion to calculating the static neutron star masses, 
noting that rotation may add as much as 0.1-0.2$\:M_\odot$ \cite{Cook}. The static 
masses are calculated by numerical integration of the Tolman-Oppenheimer-Volkoff 
equations \cite{TOV} of hydrostatical equilibrium, namely (setting G=c=1, so that all 
physical quantities are given in centimeters, $M_\odot\!\equiv\!1.477\times10^5$ cm):
\begin{eqnarray} \label{eq:TOV}
\frac{dm}{dr}=4\pi r^2\varepsilon\;\;  ; \;\;
\frac{dP}{dr}=-\frac{\varepsilon m}{r^2}
               \left(1+\frac{P}{\varepsilon}\right)\left(1+\frac{4\pi r^3P}{m}\right)
                                     \left(1-\frac{2m}{r}\right)^{-1} \\ 
\mbox{(and}\;\; 
\frac{d\Phi}{dr}=\frac{1}{\varepsilon}\frac{dP}{dr}\left(1+\frac{P}{\varepsilon}\right)^{-1}
\mbox{for the field } \Phi \mbox{)}\;\;\;       \nonumber \; .
\end{eqnarray}

These equations are solved following the recipe of Arnett and Bowers \cite{ArBow}, by 
setting a central density $\varepsilon_c$ in the star and integrating outwards until 
reaching zero pressure. The equation of state of Baym, Pethick and Sutherland \cite{BPS} 
is used for the subnuclear densities. The specific choice of the EoS for these 
densities is actually unimportant for the mass sequence, as the fraction of the 
neutron star mass originating from the subnuclear densities is insignificant. 

\subsection{Neutron Star Masses}

The predicted mass sequences for the three equations of state for nuclear matter 
$(\delta\!=\!2,\frac{5}{3}$ and $\frac{4}{3}$) are plotted in Fig. 12 as a function 
of the central density. As is expected from the large difference between these 
equations (Fig. 6), the neutron star mass sequences vary considerably from one 
equation to another. For the stiffest equation ($\delta\!=\!2$), at 
$\varepsilon_c\!=\!1.25\times10^{15}$ gm/cm$^3$, after which the EoS breaks causality, 
the mass begins to saturate at the very high value of 2.6$\:M_\odot$. The softest 
equation, which is still stiff in comparison to most of the more sophisticated nuclear 
matter models, yields a maximum mass of about 2.2$\:M_\odot$.

The consequences of the relative similarities between the equations of state for 
matter with hyperons are even enhanced in the neutron star mass calculations. As 
noted in Section 5, the dominant factor for the EoS is the number of 
species that are available for production in the matter, the case of a repulsive 
$\Sigma$-nm interaction being clearly different than the attractive one, and both 
are very different than the nuclear matter EoS.

Fig. 13 presents six plots of neutron star mass sequences. In all cases the 
hyperon-hyperon well depth is chosen as $-V_{h_1h_2}=40$ MeV, and the curves differ in 
the short range repulsive forces model, i.e. in the values of $\gamma$ and $\delta$. 
For comparison, the mass function of the softest nucler matter equation 
$\delta\!=\!\frac{4}{3}$ is also plotted. The outstanding result is that while in 
each case the masses corresponding to low densities are very model dependent in terms 
of the short range interactions (the values of $\gamma$ and $\delta$), the maximum 
masses are very similar for different equations of the same type (including or 
excluding the $\Sigma$'s). The behavior at low central densities must be different, 
as it results mainly from the nucleon-nucleon interactions, which determine the precise 
model dependence. The similarity of the mass sequences at high central densities is a 
direct consequence of the hyperon "pressure control" mechanism, discussed in the 
previous Section. This mechanism is basically dependent only on the availability of 
the various hyperon species for formation in the matter - and hence, the clear 
distinction between matter which may contain $\Sigma$'s and matter which does not.

{\em These results clearly indicate that the appearance of hyperons makes the maximum 
mass of neutron stars sensitive only to the gross properties of the baryon-baryon 
interactions}, and almost insensitive to specific details. While these results 
generated with effective equations of state can hardly be considered as proof of such 
a trend, we do believe that they do serve as indication that it is the number of 
available species that dominates the neutron star mass function at high density.

The precise values of the resulting maximum masses are also of interest. Since the 
baryonic matter EoS is softer than that for nuclear matter, the maximum neutron star 
masses are expectedly smaller. The maximum mass in the various models for matter 
without $\Sigma$'s are about 1.6-1.7$\:M_\odot$, while the softer matter with 
$\Sigma$'s holds maximal masses of only about 1.4-1.5$\:M_\odot$. The proximity of the 
latter value to the upper limit of observed neutron star masses (1.44$\:M_\odot$ of PSR 
1913+16) may be viewed in two ways: as an indication that $\Sigma$'s are bound in the 
matter, thus limiting the maximal mass to less than 1.5$\:M_\odot$; or on the contrary 
- since the EoS with $\Sigma$'s yields such a border-line result, then perhaps it 
is more plausible that $\Sigma$'s are unbound in the matter. This possibilty 
is enhanced by the recent suggestions that the mass of the Vela Pulsar is above 
1.55$\:M_\odot$, based on analysis of its X-ray source properties \cite{Vela}.

While we are unaware of an explicit indication to a hyperon induced ``pressure control''
mechanism in previous works, it is interesting to note that most studies which include 
hyperons in neutron star matter find that the maximum masses lie within a narrow range, 
typically $1.4\!-\!1.7\:M_\odot$ (see Refs. \cite{Pand71,Glenn85,newSB,Latt2,Cook,ArBow} 
and other references therein), not unlike the results presented here. On the other hand, 
published neutron star maximum masses for nuclear equations of state do indeed vary over 
a larger range ($1.5\!-\!2.5\:M_\odot$), and are usually larger than $1.8\:M_\odot$ 
[ibid.]. Both these results are consistent with the hyperon induced ``pressure control'' 
mechanism discussed above. 

We complete this study by remarking that all results show little dependence on the 
hyperon-hyperon well depth, as is expected for nucleon-dominated matter. Choosing a 
large well depth along with a steep high density repulsion ($\gamma\!=\!2$) creates 
greater repulsion at high densities, leading to a stiffer EoS and a 
slightly higher maximal mass. However, we find that this effect changes the maximum mass 
by $\Delta M\leq0.1M_\odot$ between the case of strong hyperon-hyperon interactions and 
the case when these interactions are turned off.

\subsection{Proto Neutron Star Masses}

Realisitic proto-neutron star calculations require consistent thermodynamic solutions, 
in order to establish the physical temperature profile for the star (as the star's 
physical state is probably closer to constant entropy per baryon rather than constant 
temperature). We thus limit ourselves to simplified static solutions, in order to 
recover the key feature of proto-neutron star masses.
	
Since the EoS for nuclear matter stiffens when neutrinos diffuse out of 
the proto-neutron star, the maximal mass of the fully evolved neutron star will be 
greater than that of the proto-neutron star. Hence, if a stable proto-neutron star is 
formed in the supernova (and does not collapse initially into a black hole), it will 
also have a stable configuration in the fully evovled state.

The opposite trend in matter with hyperons also brings about the possibility of the 
reverse physical scenario: since the equation of state for matter with hyperons 
softens as the neutrinos diffuse outward (and the hyperon population increases), the 
maximal mass of the proto-neutron star will be greater than that of the fully evovled 
star, under the same physical assumptions. Thus, the stage is set for a delayed 
collpase of the proto-neutron star into a black hole, if its mass will be too large 
for the neutrino-free equation of state. Such a scenario has been motivated by the 
combination of a neutrino pulse in SN1987A (so there was no initial collapse into a 
black hole), and the fact that no pulsar has been detected; see Refs. [7-10].

We reproduced the necessary features described above in the effective EoS 
scheme, when deriving mass sequences for proto-neutron stars and evolved neutron 
stars (Fig. 11). For nuclear matter the EoS for proto-neutron star (neutrino-trapped) 
is softer than for evolved neutron star (neutrino-free) and the maximal mass drops 
from $\sim2.4\:M_\odot$ to $\sim2.0\:M_\odot$ for the proto-neutron star. Since the 
EoS for neutrino-trapped matter is stiffer for matter with hyperons, the maximal mass 
of the proto-neutron star is found to be larger for the hyperonic case. For matter 
with all types of hyperons the maximum mass of the proto-neutron star is almost 
$1.7\:M_\odot$ (with respect to a little over $1.4\:M_\odot$ for the evolved neutron 
star). Matter with no $\Sigma$'s is less affected by the neutrino diffusion (as 
explained above, the $\Sigma$'s are somewhat suppressed in neutrino-trapped matter) 
but the proto-neutron star still has a maximum mass of about $0.1\:M_\odot$ higher than 
that of the evovled star ($1.8\:M_\odot$ vs. $1.7\:M_\odot$). This trend appears for 
the other various cases of assumptions made about the baryon-baryon 
interaction and shows only little model dependence, as in the case of the evolved 
neutron star masses. 

\section{Conclusions and Discussion}

The effective equation of state described in the present study is designed to allow 
rapid calculations and extensive parameter surveys of high density matter rich with 
strange baryons. This equation generalizes the Lattimer-Swesty EoS for nuclear matter, 
and is constructed so that it will reproduce the physical features attributed to such 
matter, while not presuming to correctly describe the microscopic details of the 
underlying physics. Its construction is performed by basing the EoS on local 
(effective) potentials for the various componenets of the baryon-baryon interaction, 
which are constrained as much as possible by experimental data from hypernuclei.

In view of the large uncertainties in the baryon-baryon interactions, especially at 
high densities, we examine the properties of the EoS with various assumptions 
regarding these interactions. We investigate sensitivity to the density dependence of 
the repulsive term and remark on the unknown power of the hyperon-hyperon 
interactions. Special attention is given to the consequences of a recent analysis of 
$\Sigma$-atoms data, which suggests a strong isoscalar repulsion in the 
$\Sigma$-nuclear matter ($\Sigma$-nm) interaction (very different than the commonly 
assumed interaction), and excludes $\Sigma$ hyperons from forming in the high density 
matter.

We find that the effective equation of state yields similar equilibrium compositions 
and equations of state as found in equivalent RMF models. 
The main features of the hyperon formation in zero-temperature neutrino free matter 
(typical of evolved neutron star interiors) are:

\begin{enumerate}
\setlength{\parskip}{0.0in}
\item Hyperons begin to appear in the matter at about twice the nuclear saturation 
density $\rho_s\!\approx\!0.155\,$ fm$^{-3}$, and at $2.5\rho_s$ the hyperon fraction 
in the matter is already significant. First to appear are the $\Sigma^-$ (due 
to its negative charge) and the $\Lambda$, or the $\Lambda$ alone, depending on the 
assumed $\Sigma$-nm interaction. Other hyperons follow at higher densities.

\setlength{\parskip}{0.0in}
\item Hyperon production is followed by considerable deleptonization, as 
negatively charged hyperons replace electrons and muons in maintaining charge 
neutrality.
\end{enumerate}
\setlength{\parskip}{0.1in}

We find that these features are typical of all models, although the details are 
model-dependent as far as the density dependence of the baryon-baryon repulsive interaction 
terms. Since the matter is dominated by nucleons in most of the density range in question, 
there is little dependence on the hyperon-hyperon interactions.	

The key effect of the appearance of hyperons in the matter is the softening of the 
equation of state, due to the lowering of both repulsion and Fermi energies as the 
total baryon density is divided among a larger number of species. We emphasize that 
{\em hyperons serve as a ``pressure control'' mechanism in the high density 
matter}, since the extent of their appearance is directly related to the details of 
the nuclear interactions. Thus, while for nuclear matter a more powerful 
nucleon-nucleon repulsion creates a stiffer EoS, in baryonic matter an enhanced 
production of hyperons will occur, restoring the gross properties of the EoS. The 
immediate result of this ``pressure control'' by the hyperons is that all EoS 
including hyperons are very similar, the only significant model dependence being on 
the number of available species (i.e. the general features of the $\Sigma$-nm 
interaction discussed above).

The most notable manifestation of this effect is in the predicted maximum masses of 
neutron stars. By solving the static mass functions of neutron stars for the various 
equations, we find that the hyperon-imposed similarity in the equations of state 
leads to very moderate specific-model-dependence of the maximum masses. {\em While 
the maximum mass for nuclear matter is highly model dependent, the maximum masses for 
matter with hyperons depend almost entirely just on the number of available 
species}. Specifically, we find that for matter which includes $\Sigma$'s, the 
maximum mass for all equations of state are about $1.4\!-\!1.5\:M_\odot$, while for the 
case of excluded $\Sigma$'s, maximum masses are all in the range $1.6\!-\!1.7\:M_\odot$. 
Such analysis with an effective equation of state may hardly be considered as a proof 
of this point, but we do believe that the general trend described above indicates a 
genuine physical principle.

It is interesting that the result for matter with $\Sigma$'s yields just about the 
maximum observed neutron star mass (the small difference can be bridged by allowing 
also for rotation and finite temperature). This may be viewed in two ways: one could 
either argue that this is too border-line a result, implying $\Sigma$'s should be 
excluded, or rather find this an indication that the maximum mass is indeed close to 
the observed value, as determined by the hyperons. Clearly further analysis is needed 
to elaborate on this point.	

We briefly investigate the properties of proto-neutron star matter, which has a 
finite temperature and in which neutrinos are assumed to be trapped, so that the 
total lepton number remains constant. Such matter is relevant to the simulation of 
proto-neutron star evolution and core-collapse process in supernovae. Again we find 
that hyperons appear at about $2\rho_s$ and accumulate significantly by $2.5\rho_s$, 
with very weak model dependence. Hyperon appearance changes the trend of the neutrino 
population, which  begins to increase rather than decrease at higher 
densities, since the hyperons indirectly lower the electron fraction required to 
maintain charge neutrality. These results are in good agreement with studies 
performed with more sophisticated models. 

The formation of hyperons softens the equation of state for core-collapse matter as 
well, and once again serves as ``pressure control'' for the nucleons. The softening 
occurs, of course, at the density of hyperon accumulation, which is border-line with 
regard to the densities reached during the supernova. None the less, a softer 
equation of state is appealing, as it will strengthen the shock and the neutrino 
delayed shock, thus strenghtening the explosion. In the proto-neutron star case such 
a softening is clearly important, and we do reproduce the well-known tendency of the 
maximum mass for matter with trapped neutrinos being larger than that for neutrino 
free matter. This feature, which is typical for matter with hyperons and opposite to 
the one prevalent in nuclear matter, opens the way for the ``delayed collapse'' 
scenario, motivated by the lack of an observed pulsar in the Supernova 1987A remnant. 
In such a scenario, a newly formed neutron star could exist in a meta-stable 
configuration, collapsing into a black hole when the neutrinos diffuse out of the 
star leaving it with too large a mass.

All of the above results serve as indication that the effective equation of 
state may be treated as a trustworthy tool in the analysis of high density matter 
with strangeness. Nonetheless, some further extension of the equation should be 
considered. One area for improvement is the description of the high density behavior 
of the forces. A possible approach would be to add another repulsive term 
$c'\rho^{\gamma'}$, where $\gamma'$ will be greater than $\gamma$ (which could then 
be set to a low value allowing consistently for a low nuclear incompressibility). 
This new repulsive term could be set universaly for all species (e.g, if it may be 
attributed to quark degrees of freedom). This is somewhat similar to a suggestion by 
Lattimer, Swesty and Myra \cite{SwesLattMy}, that the repulsive term in the energy density 
formula be amended to $c\rho^{\delta+1}/(1+d\rho^{\delta-1})$, as an ad hoc method to 
artificially soften the equation.

Other possible improvements are introducing finite ranges to the forces, and a more 
comprehensive treatment of effective masses. Clearly, any future experimental and 
theoretical constraints on the nuclear, nucleon-hyperon and hyperon-hyperon forces 
must be incorporated into the equation and its parameters as well.

Introduction of $\pi$ and K condensates should also be considered, although viewed as 
unlikely by several current works. The same applies to the formation of $\Delta$ 
isobars, which are found to be excluded from the matter by using standard assumptions 
regarding its interactions and mass. A natural extension of the equation would be to 
include a deconfined quark phase at high densities, for which an effective equation of 
state is quite simple \cite{AlFarOlSS}.

Further work is also required regarding the implications of strangeness on high 
density matter in astrophysical contexts. Up to date no neutron star property that 
can be uniquely connected to hyperon formation has been suggested, although observed 
cooling rates have been shown to place some limits on the matter composition 
(a large hyperon fraction along with a significant lepton population would produce 
much too large cooling rates through the direct Urca mechanism). Studies of possible 
implications of hyperon formation on rotational properties of neutron stars, along 
with magnetic, superconducting and superfluid behavior of such matter, and crust 
properties with relation to observed glitches should also be considered (far more 
extensive studies on these subjects have been carried out for strange-quark-stars, 
following the still debated stable-strange-matter hypothesis). We note that a recent 
work by Glendenning, Pei and Weber \cite{GPWbrkind} suggests that the existence of a 
quark-matter core (which will definitely include a non-zero-strangeness-fraction) could 
be identified by its effect on the pulsar breaking index, if such a core builds up 
during the pulsar spin down.  

A unique signature of hyperon (and generally strange particles) production in 
proto-neutron star would also be of considerable interest, especially regarding the 
``delayed-collapse'' scenario and whether it may lead to an appreciable effect on the 
neutrino pulse from the proto-neutron star \cite{BethBrn}. Studies along these lines 
have begun in recent years, and are well worth continuing.

Last but clearly not least, the results presented in 
this study imply that hyperon formation may be of importance in core-collapse 
supernovae, if densities equal or greater to twice the nuclear saturation density are 
reached in the collapsing core prior to the ``bounce''. A softened equation of state 
might claim its role in the shock-wave/neutrino-delayed-shock process, presumably 
strengthening the explosion.	
\setlength{\parskip}{0.5in}

We are grateful to Sidney Kahana for stimulating discussions about high 
density matter and baryon-baryon interactions. We further wish to thank 
Eliahu Friedman, Itamar Lichtenstadt, Greg Cook and Nir Barnea for helpful 
suggestions and advice during the preparation of this manuscript. This research was 
supported in part by the U.S-Israel Binational Science Foundation.

\pagebreak

\pagebreak

\begin{center}
\large{Figure Captions}
\end{center}
\setlength{\parskip}{0.1in}

Fig. 1. The density dependence of $\Lambda$ potential in 
nuclear matter, $V_{\Lambda N}(\rho_N)=a_{\Lambda N}\rho_N+c_{\Lambda N}\rho_N^\gamma$,
for different values of the exponent $\gamma$ in the repulsion term, using the 
parametrs sets of MDG \cite{MDG} given in Table 2. The arrow marks the nuclear 
saturation density, $\rho_s\!=\!0.155$ fm$^{-3}$.

Fig. 2. The equilibrium compositions of nuclear matter (n-p-e-$\mu$) as 
a function of baryon density for the case $\delta=2$; see Eq. (\ref{eq:epotNN}) and 
Table 1. Other models of the nuclear interaction (different values of $\delta$) yield 
similar equilibrium compositions.

Fig. 3a. The equilibrium compositions for matter containing hyperons as 
well as nucleons and leptons, for $\delta\!=\!\gamma\!=\!2$ (see 
Eq. (\ref{eq:epotbar})), and for the hyperon-hyperon well depth $-V_{h_1h_2}=40$ MeV.

Fig. 3b. Same as 3a but with $\delta\!=\!\gamma\!=\!\frac{5}{3}$.

Fig. 3c. Same as 3a but with $\delta\!=\!\gamma\!=\!\frac{4}{3}$.

Fig. 4a. The equilibrium compositions for matter containing $\Lambda$'s
and $\Xi$'s as well as nucleons and leptons, but no $\Sigma$'s (when a repulsive 
$\Sigma$-nm interaction is assumed), for $\delta\!=\!\gamma\!=\!2$, 
and a hyperon-hyperon well depth $-V_{h_1h_2}=40$ MeV.

Fig. 4b. Same as 4a but with $\delta\!=\!\gamma\!=\!\frac{5}{3}$.

Fig. 4c. Same as 4a but with $\delta\!=\!\gamma\!=\!\frac{4}{3}$.

Fig. 5. Same as 3a but with hyperon-hyperon interactions ``turned off''.

Fig. 6. Equations of state for nuclear matter (n-p-e-$\mu$) for the 
different models  of the nucleon potential energy density. The 
equations for $\delta\!=\!2$ and $\frac{5}{3}$ break causality at 
$1.25\times10^{15}$ gm/cm$^3$ and $1.94\times10^{15}$ gm/cm$^3$, respectively, as 
indicated by the dashed curves.

Fig. 7. Equations of state for high density matter with nucleons and 
leptons alone; nucleons, hyperons and leptons; and $\Lambda$'s
and $\Xi$'s as well as nucleons and leptons, but no $\Sigma$'s (for a repulsive 
$\Sigma$-nm interaction). The parameters used are $\delta\!=\!\gamma\!=\frac{5}{3}$, 
and $-V_{h_1h_2}=40$ MeV.

Fig. 8. Equations of state for high density matter with nucleons, 
hyperons (assuming an attractive $\Sigma$-nm interaction) and leptons, for different 
choices of the baryon potential energy density dependence, 
$\delta\!=\!\gamma\!=\!2,\frac{5}{3}$ and $\frac{4}{3}$, and a hyperon-hyperon well 
depth $-V_{h_1h_2}=40$ MeV. The equations of state correspond to the 
equilibrium compositions of Figs. 3a, 3b and 3c.

Fig. 9. The effect of hyperon formation on the adiabatic index $\Gamma$ 
of the EoS for the case of $\delta\!=\!\gamma\!=\!\frac{5}{3}$, and 
$-V_{h_1h_2}=40$ MeV, compared to the corresponding EoS for nuclear matter. Each 
``dip'' in the curve of the adiabatic index corresponds to the appearence of a new 
hyperon species.

Fig. 10a.  The equilibrium compositions of neutrino trapped nuclear 
matter (n-p-e-$\nu_e$) as a function of baryon density for the case 
$\delta\!=\!\frac{5}{3}$, with a lepton fraction of $Y_l=0.36$ and temperature 
$T\!=\!20$ MeV. The equilibrium compositions for other models of the nuclear 
interaction (values of $\delta$) are very similar.

Fig. 10b. The equilibrium compositions for neutrino-trapped matter 
containing hyperons as well as nucleons and leptons, where $\delta\!=\!\gamma\!
=\!\frac{5}{3}$, and $-V_{h_1h_2}=40$ MeV.

Fig. 10c. The equilibrium compositions for neutrino-trapped matter 
containing hyperons as well as nucleons and leptons, but no $\Sigma$'s (for a 
repulsive $\Sigma$-nm interaction), for $\delta\!=\!\gamma\!=\!\frac{5}{3}$ and 
$-V_{h_1h_2}=40$ MeV.

Fig. 11. Equations of state for neutrino-free matter (solid lines) and 
neutrino-trapped matter (dahed lines) for $\delta\!=\!\gamma\!=\!\frac{5}{3}$, for the 
different versions of high density matter: nucleons only (a), nucleons and all hyperons 
(b) and nucleons, $\Lambda$'s and $\Xi$'s but no $\Sigma$'s (c). It is noticeable that 
the EoS for the neutrino-trapped case is softer for nuclear matter, but stiffer for 
matter with hyperons.

Fig. 12. Neutron star mass vs. central density mass-sequences for 
nuclear matter (n-p-e-$\mu$) for the different models of the nucleon potential energy 
density.

Fig. 13. Neutron stars mass vs. central density mass-sequences for high 
density matter with all hyperon species and with only $\Lambda$'s and $\Xi$'s but no 
$\Sigma$'s (for a repulsive $\Sigma$-nm interaction), for different 
values of $\gamma$ and $\delta$. In all cases the hyperon-hyperon well depth is  
$-V_{h_1h_2}=40$ MeV. Also plotted is the neutron star mass sequence for the softest 
equation for nuclear matter $\delta\!=\!\frac{4}{3}$.

\pagebreak

\setlength{\parskip}{0.1in}
\begin{table} 
\caption{\label{tb:nonvals}} Values for the parameters of the terms corresponding to 
the nuclear potential energy density of Eq. (\ref{eq:epotNN}), as derived from the 
Lattimer-Swesty equation of state with $m^*\!=\!m$.
\begin{center}
\begin{tabular}{c c c c} \hline
$\delta$       & $a_{NN}$                          & $b_{NN}$             
               & $c_{NN}$                                   \\
               & [MeV fm$^3]$   & [MeV fm$^3]$
               & [MeV fm$^{3\delta}]$                   \\ \hline
$ 2 $          & $-784.4$  & $214.2$  & $1936.0$                 \\
$\frac{5}{3}$  & $-935.4$  & $214.2$  & $1557.2$                 \\
$\frac{4}{3}$  & $-1384.6$ & $214.2$  & $1672.8$                 \\ \hline
\end{tabular}
\end{center}
\end{table}
\begin{table} 
\caption{\label{tb:lonvals}} Values for the parameters of the terms corresponding to 
the $\Lambda$-nm interactions in the local potential energy model of Eq. 
(\ref{eq:epotbar}), for a well-depth of $-V_{\Lambda N}\!=\!28$ MeV at nuclear 
saturation density.               
\begin{center}
\begin{tabular}{c c c} \hline
$\gamma$       & $a_{\Lambda N}$ & $c_{\Lambda N}$ \\ 
               & [MeV fm$^{3}]$ & [MeV fm$^{3\gamma}]$ \\ \hline
$ 2 $          & $-340.0$ & $1087.5$                 \\ 
$\frac{5}{3}$  & $-387.0$ & $738.8$                  \\ 
$\frac{4}{3}$  & $-505.2$ & $605.5$                  \\ \hline
\end{tabular}
\end{center}
\end{table}
\begin{table} 
\caption{\label{tb:sigvals1}} Values for the parameters of the terms corresponding to 
the $\Sigma$-nm interactions in the local potential energy model  of Eq. 
(\ref{eq:epotbar}), for a well-depth of $-V_{\Sigma N}\!=\!28$ MeV at nuclear 
saturation density, assuming a $\Sigma$-nm interaction equal to the $\Lambda\!
$-nm one, except for a nonzero isovector term twice that of the N-nm interaction.
\begin{center}
\begin{tabular}{c c c c} \hline
$\gamma$       & $a_{\Sigma N}$     &  $b_{\Sigma N}$ 
               & $c_{\Sigma N}$  \\ 
               & [MeV fm$^{3}$] &  [MeV fm$^{3}$]
               & [MeV fm$^{3\gamma}]$ \\ \hline 
$ 2 $          & $-340.0$ &  $214.2$  &  $1087.5$                \\
$\frac{5}{3}$  & $-387.0$ &  $214.2$  &  $738.8$                 \\ 
$\frac{4}{3}$  & $-505.2$ &  $214.2$  &  $605.5$                 \\ \hline
\end{tabular}
\end{center}
\end{table}
\begin{table} 
\caption{\label{tb:sigvals2}} Values for the parameters of the terms corresponding to 
the $\Sigma$-nm interactions in the local potential energy model of Eq. 
(\ref{eq:signopt}), accepting the strong $\Sigma$-nm isoscalar repulsion argued 
for in Ref. \cite{BattFrGal}. The $b$ and $B$ parameters are given in fm.          
\begin{center}
\begin{tabular}{c c c c c} \hline
$\alpha$       & $b_0$  & $B_0$   & $b_1$   &  $B_1$              \\ \hline
$ 0.5  $       &  2.1   & $-4.3$  & $-1.0$  &    0                \\ \hline
\end{tabular}
\end{center}
\end{table}
\begin{table} 
\caption{\label{tb:convals1}} Values for the parameters of the terms corresponding to 
the $\Xi$-nm interactions in the local potential energy model of Eq. 
(\ref{eq:epotbar}), for a well-depth of $-V_{\Xi N}\!=\!24$ MeV at nuclear saturation 
density.               
\begin{center}
\begin{tabular}{c c c c} \hline
$\gamma$       & $a_{\Xi N}$       & $b_{\Xi N}$ &  $c_{\Xi N}$  \\
               & [MeV fm$^{3}]$   & [MeV fm$^{3}]$
               & [MeV fm$^{3\gamma}]$                  \\ \hline
$ 2 $          & $-291.5$  &  $0$  & $932.5$                  \\ 
$\frac{5}{3}$  & $-331.8$  &  $0$  & $663.4$                  \\ 
$\frac{4}{3}$  & $-434.4$  &  $0$  & $520.1$                  \\ \hline
\end{tabular}
\end{center}
\end{table}
\begin{table}
\caption{\label{tb:YYvals}} Values for the parameters of the terms corresponding to 
the various hyperon-hyperon interactions in the local potential energy model of Eq. 
(\ref{eq:epotbar}), assuming the hyperon-hyperon well-depth at a density equal to the 
nuclear saturatrion density to be $-V_{h_1h_2}\!=\!40$ MeV. Only the $\Sigma$'s are 
assumed to have a nonzero isovector component in their interactions.
\begin{center}
\begin{tabular}{c c c c c}  \hline
$\gamma$  
& $a_{YY}$ & $b_{\Lambda Y}\!\!=\!\!b_{\Xi Y}$ & $b_{\Sigma\Sigma}$ & $c_{YY}$ \\
    & [MeV fm$^{3}$] & [MeV fm$^{3}$] 
    & [MeV fm$^{3}$] & [MeV fm$^{3\gamma}$]
\\ \hline
$ 2 $          & $-486.2$ &  0  &  $428.4$ & $1553.6$  \\ 
$\frac{5}{3}$  & $-552.6$ &  0  &  $428.4$ & $1055.4$  \\ 
$\frac{4}{3}$  & $-723.2$ &  0  &  $428.4$ & $869.0$  \\ \hline
\end{tabular}
\end{center}
\end{table}

$    $
\newpage

$    $
\newpage

\epsffile{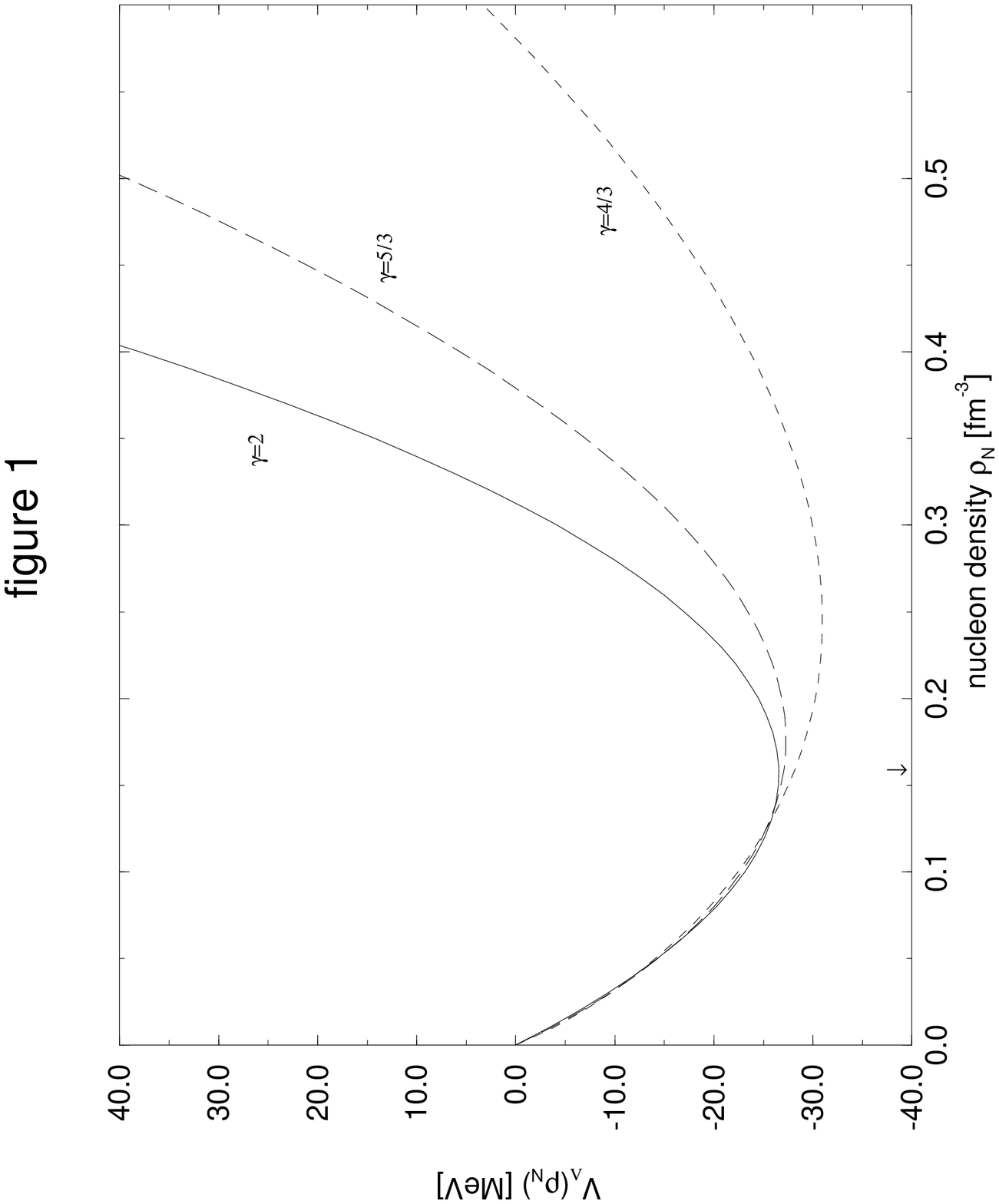}
\newpage
\epsffile{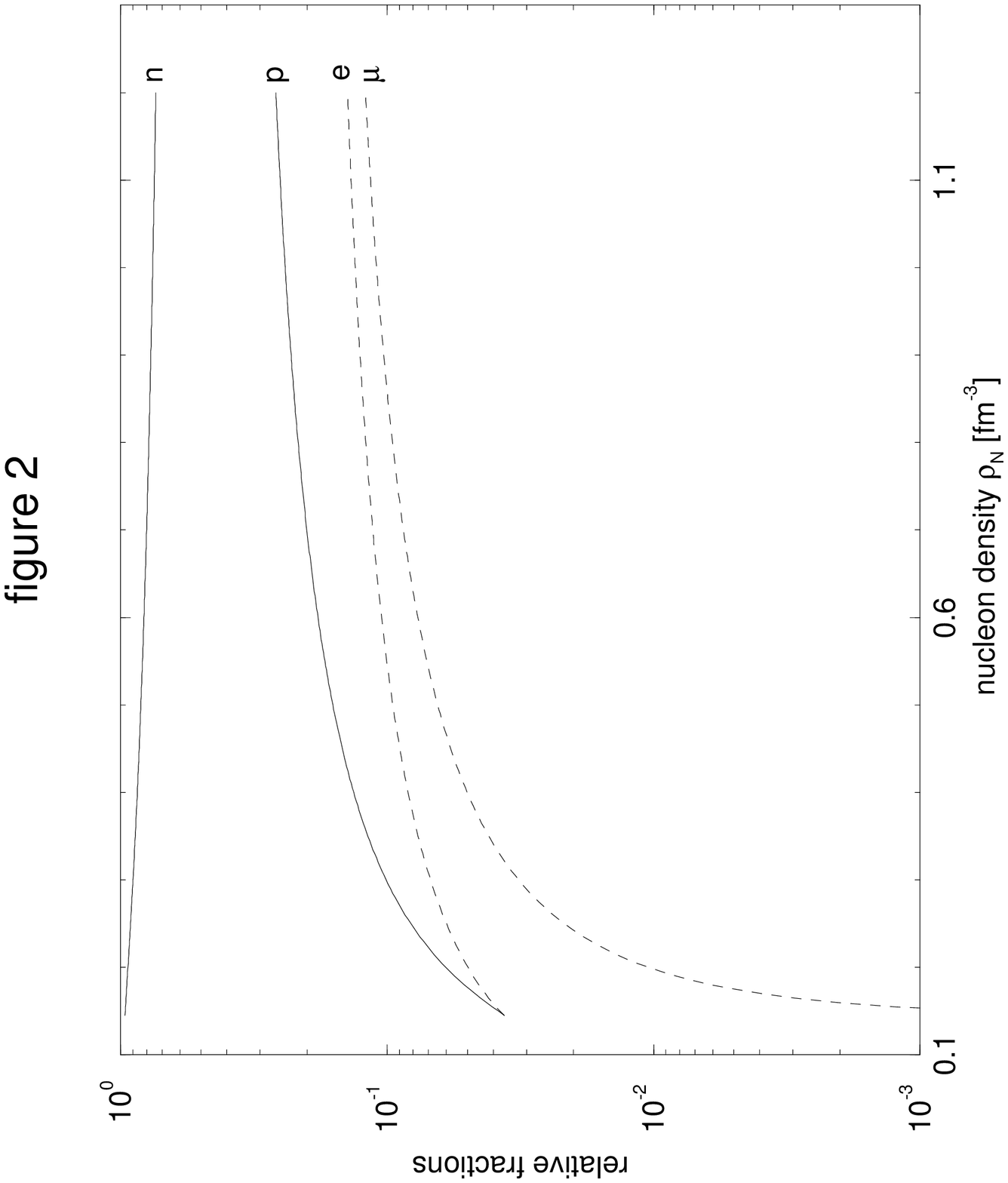}
\newpage
\epsffile{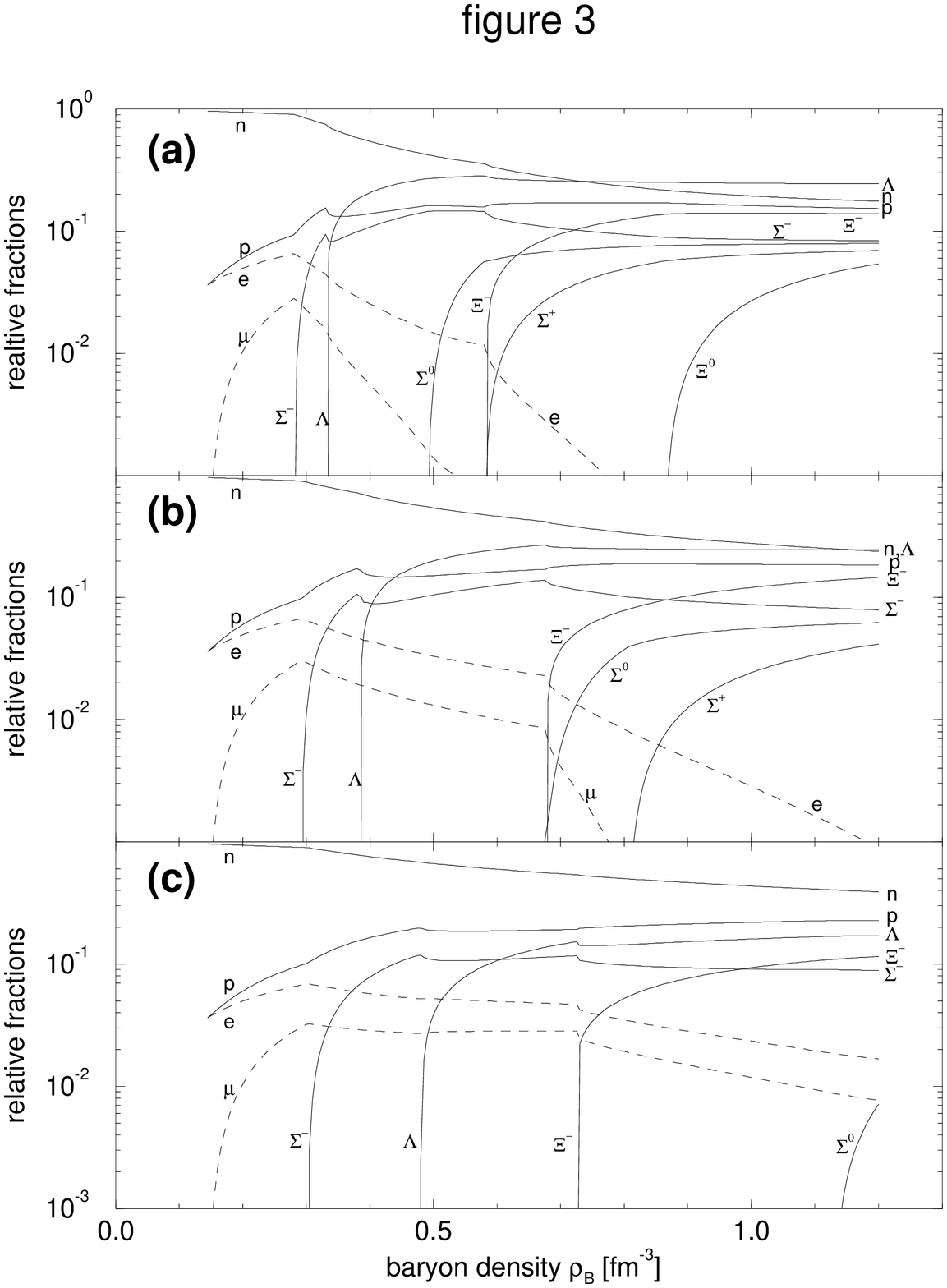}
\newpage
\epsffile{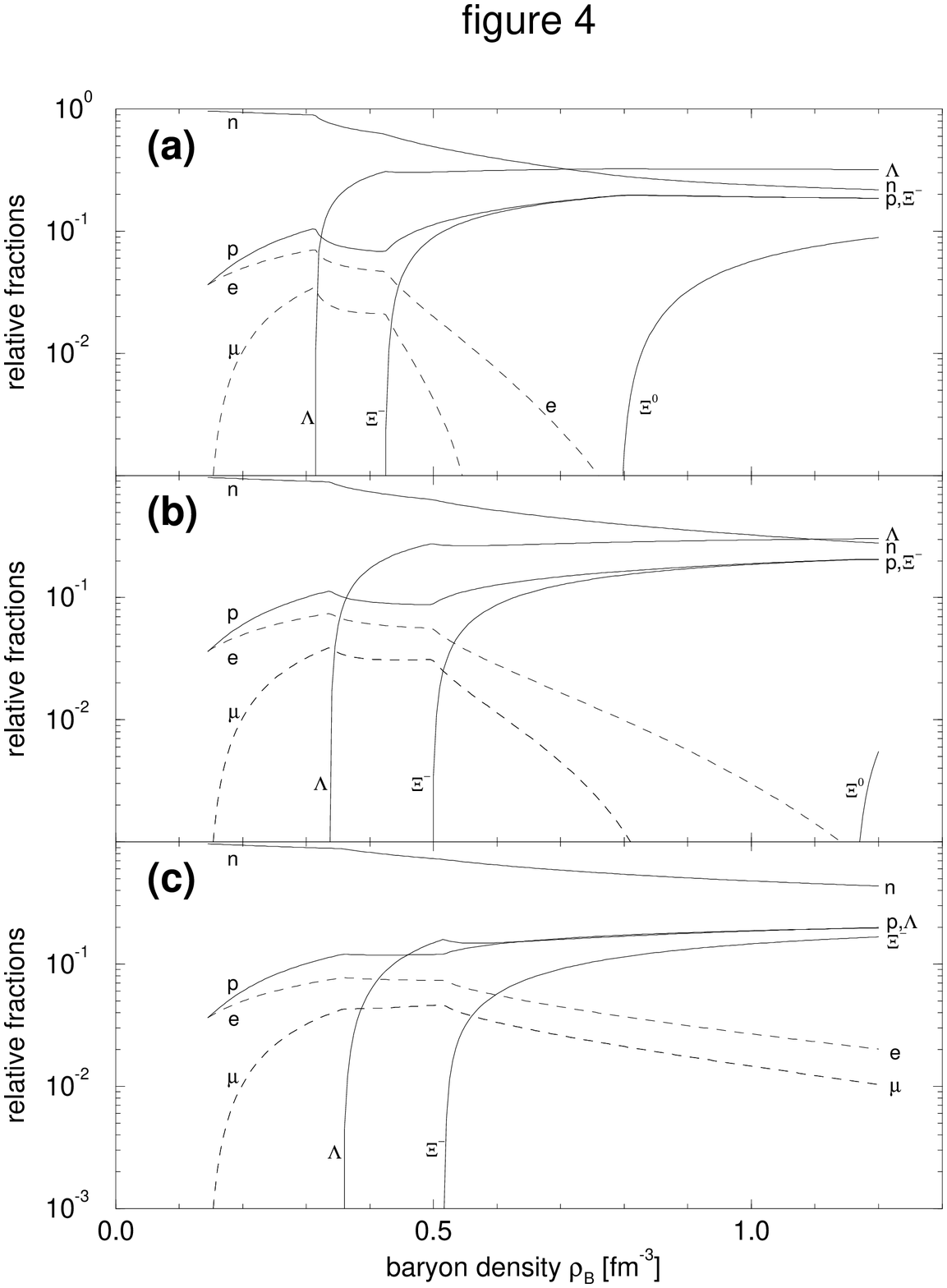}
\newpage
\epsffile{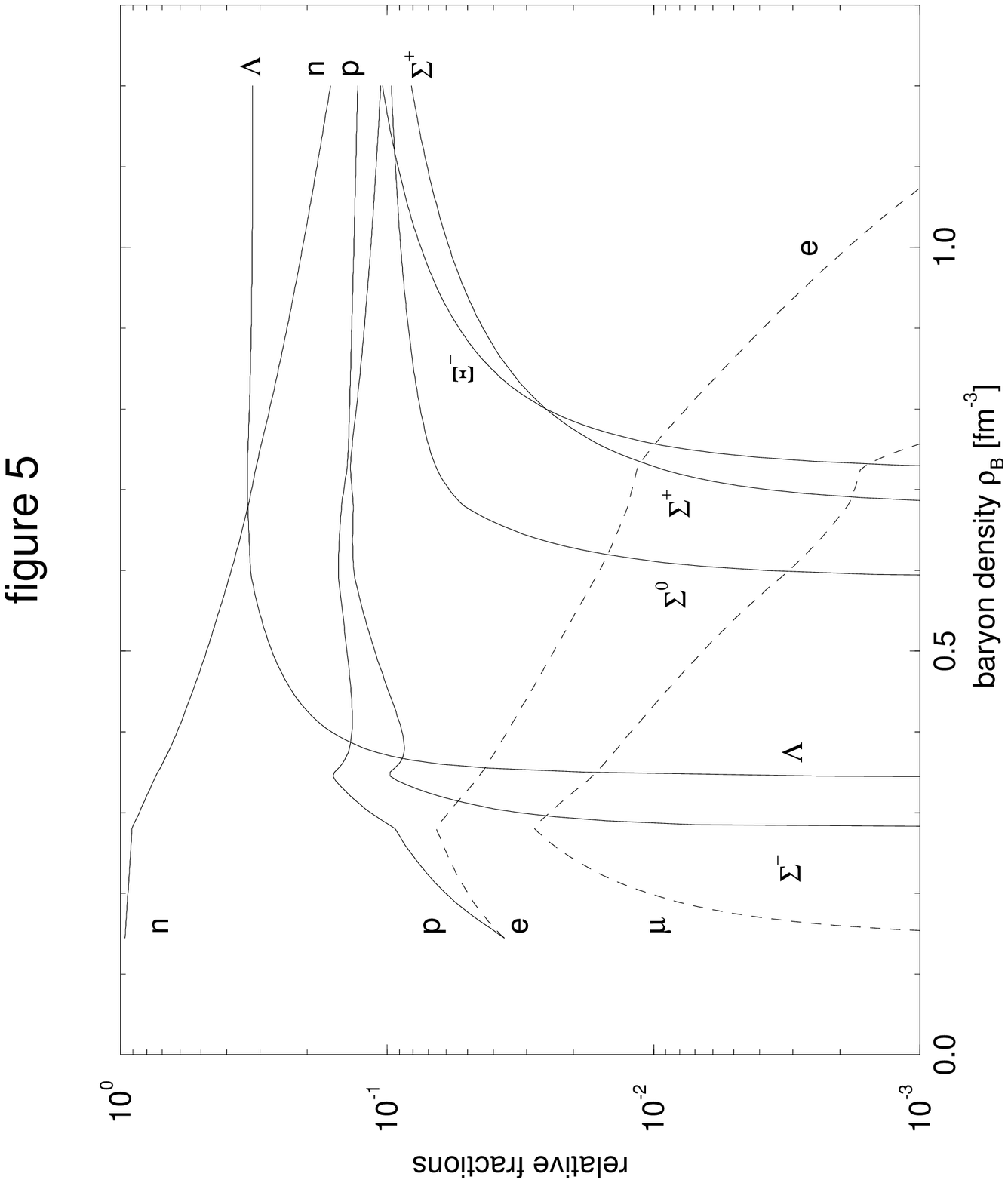}
\newpage
\epsffile{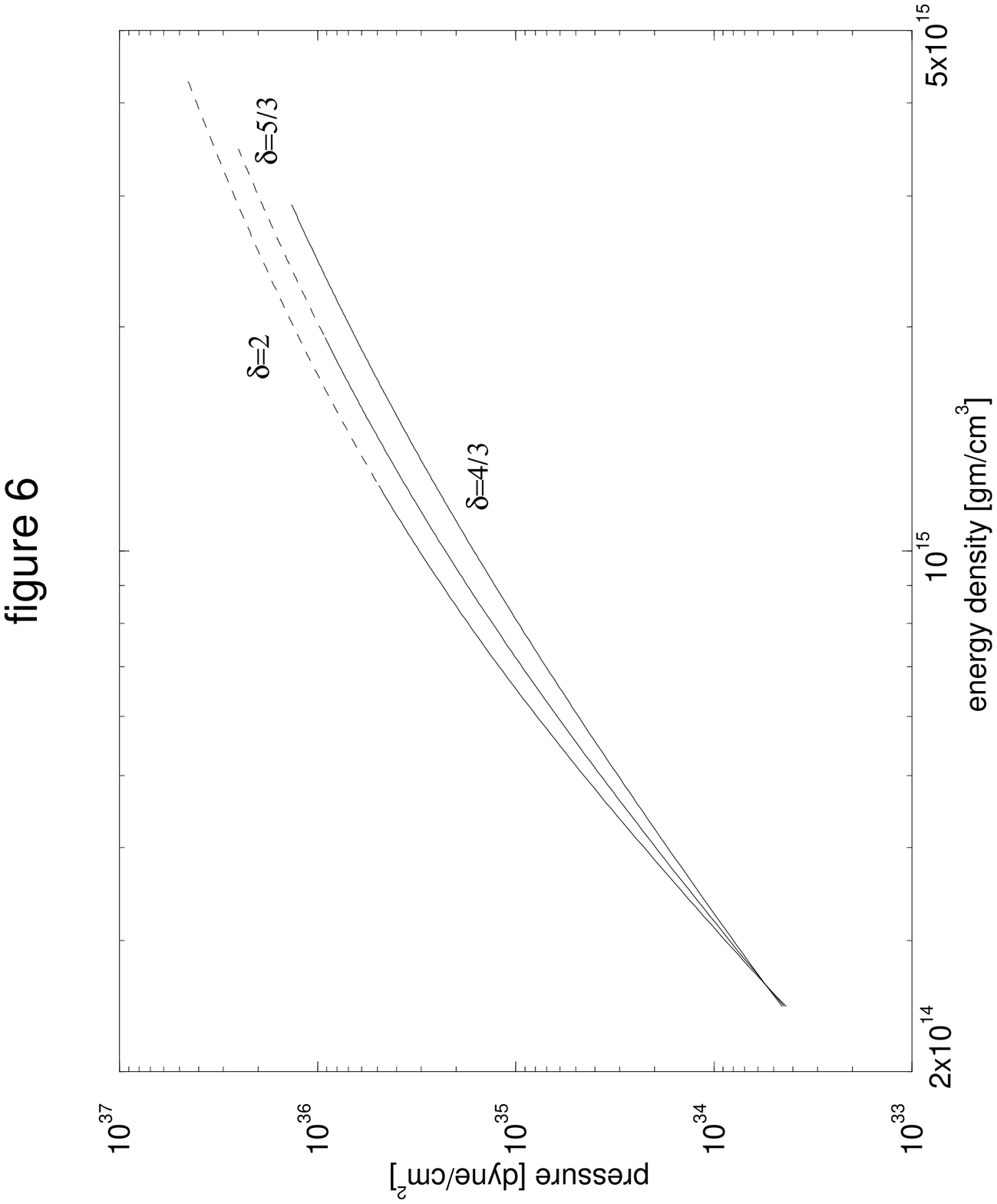}
\newpage
\epsffile{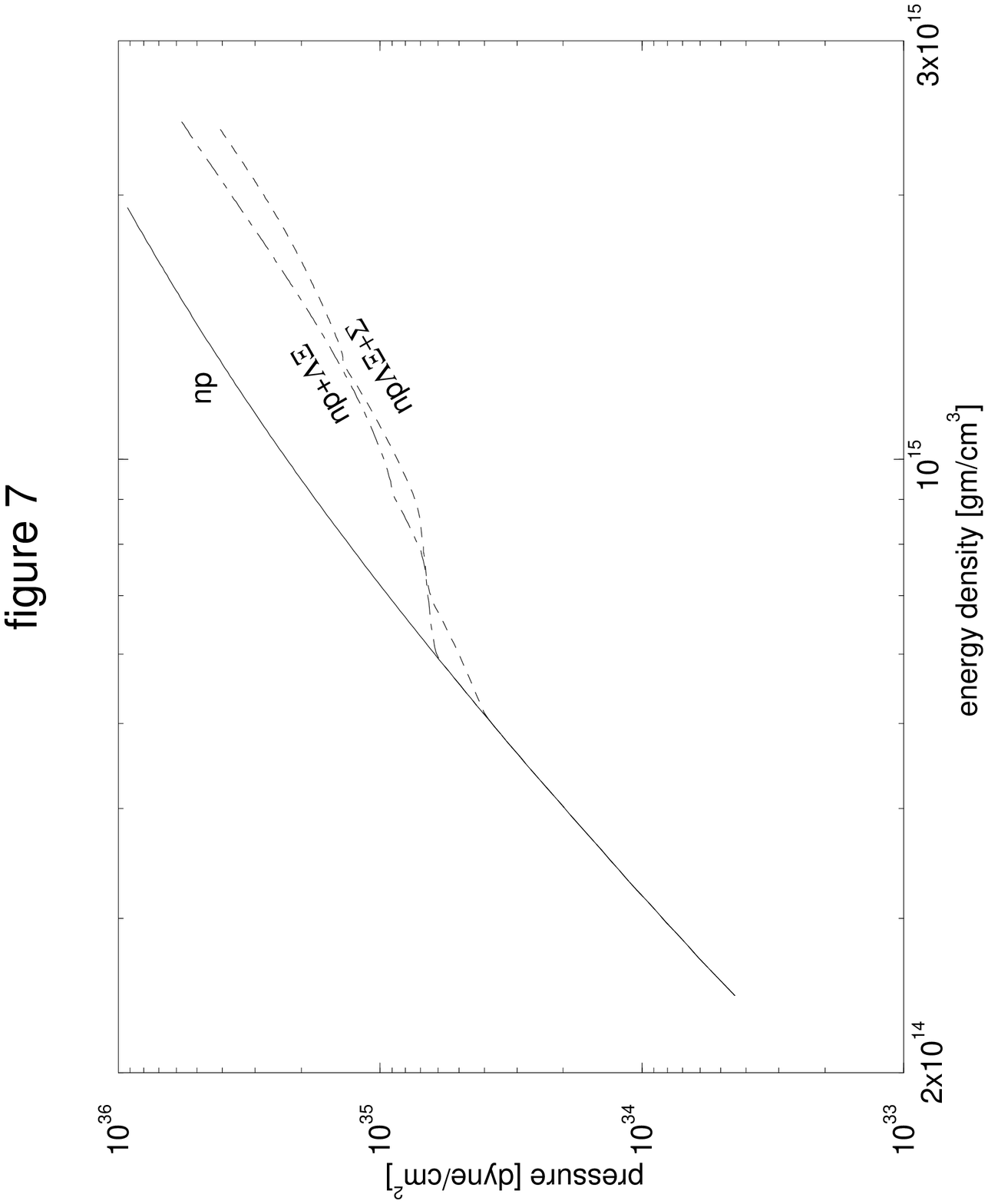}
\newpage
\epsffile{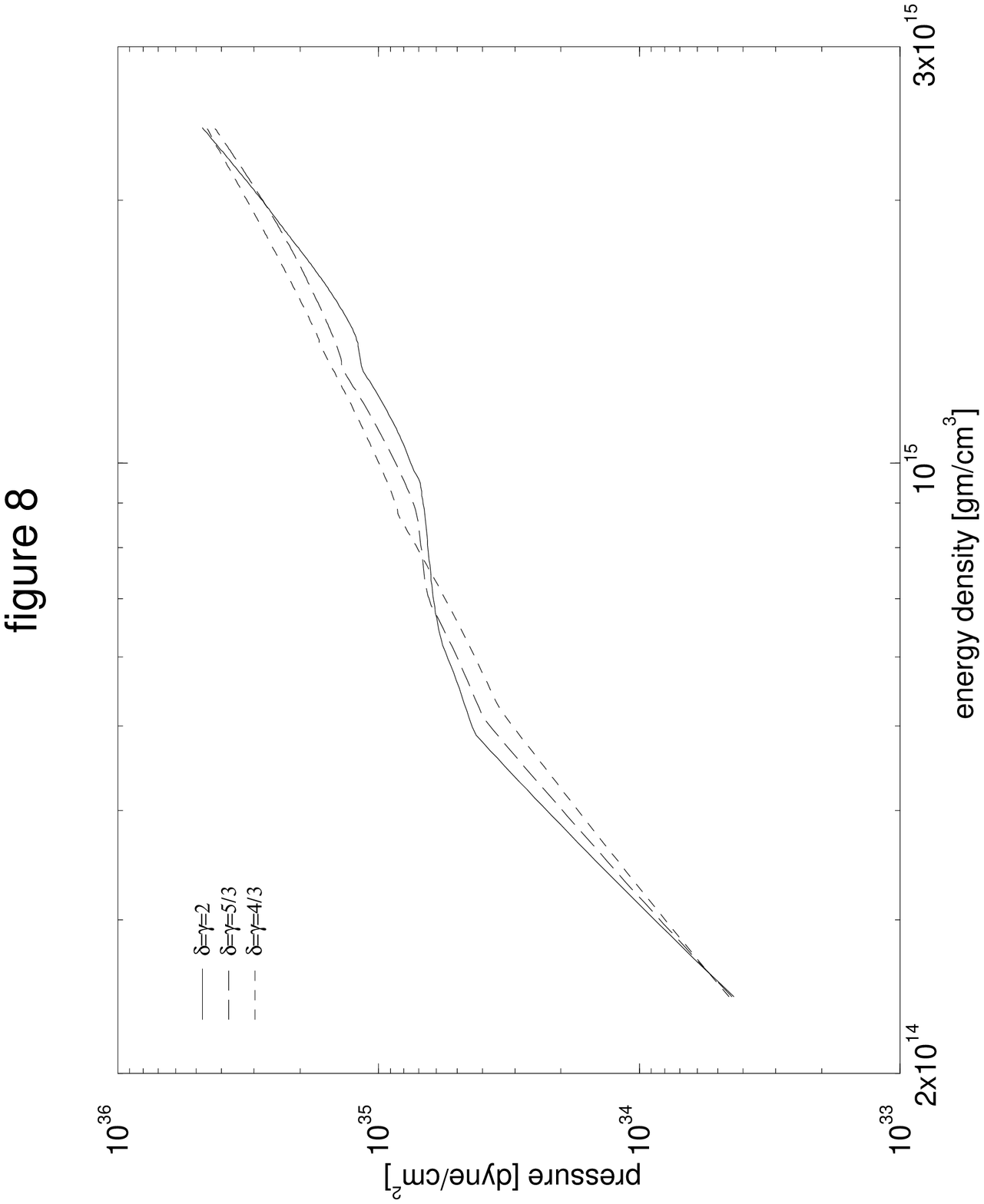}
\newpage
\epsffile{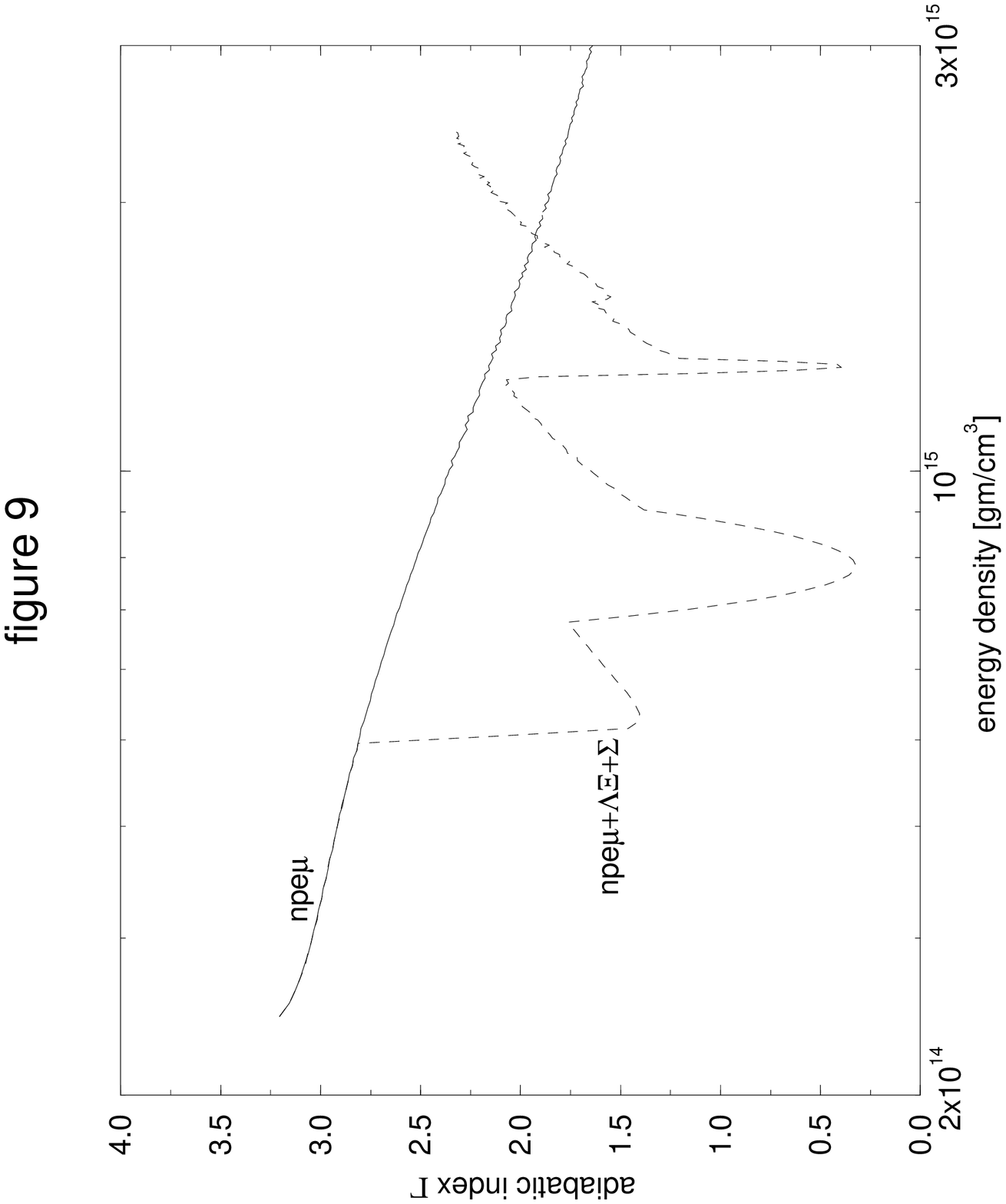}
\newpage
\epsffile{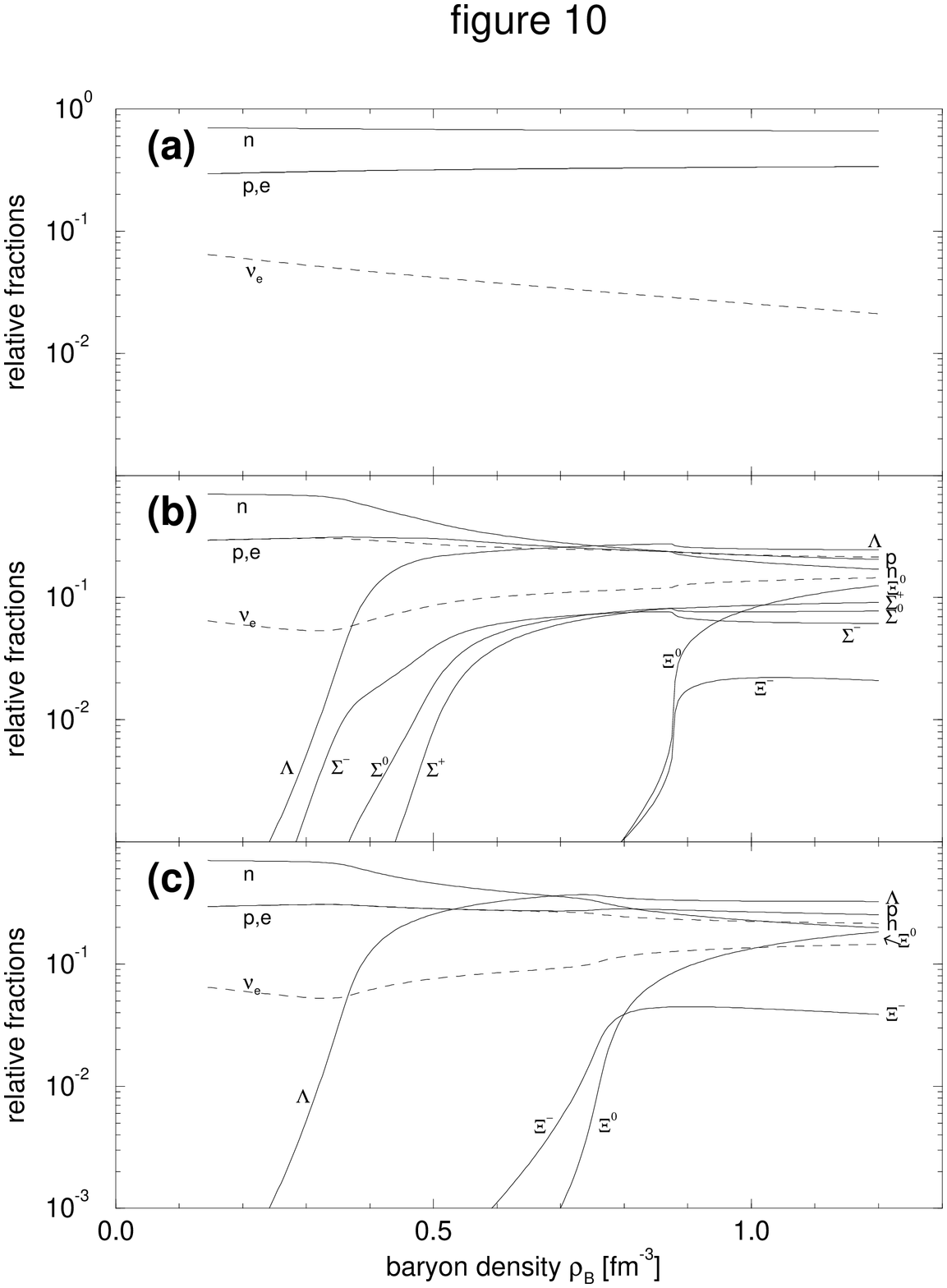}
\newpage
\epsffile{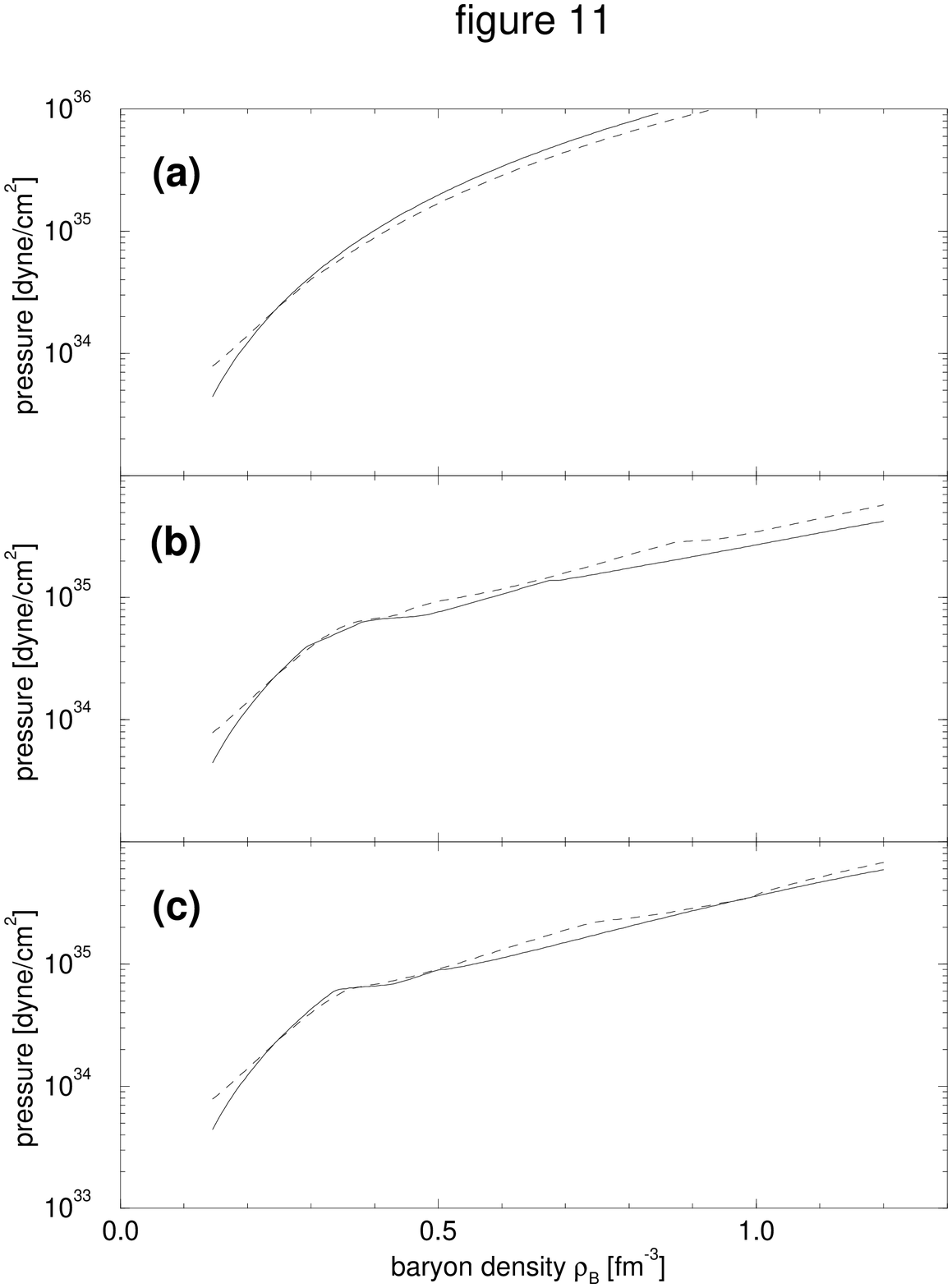}
\newpage
\epsffile{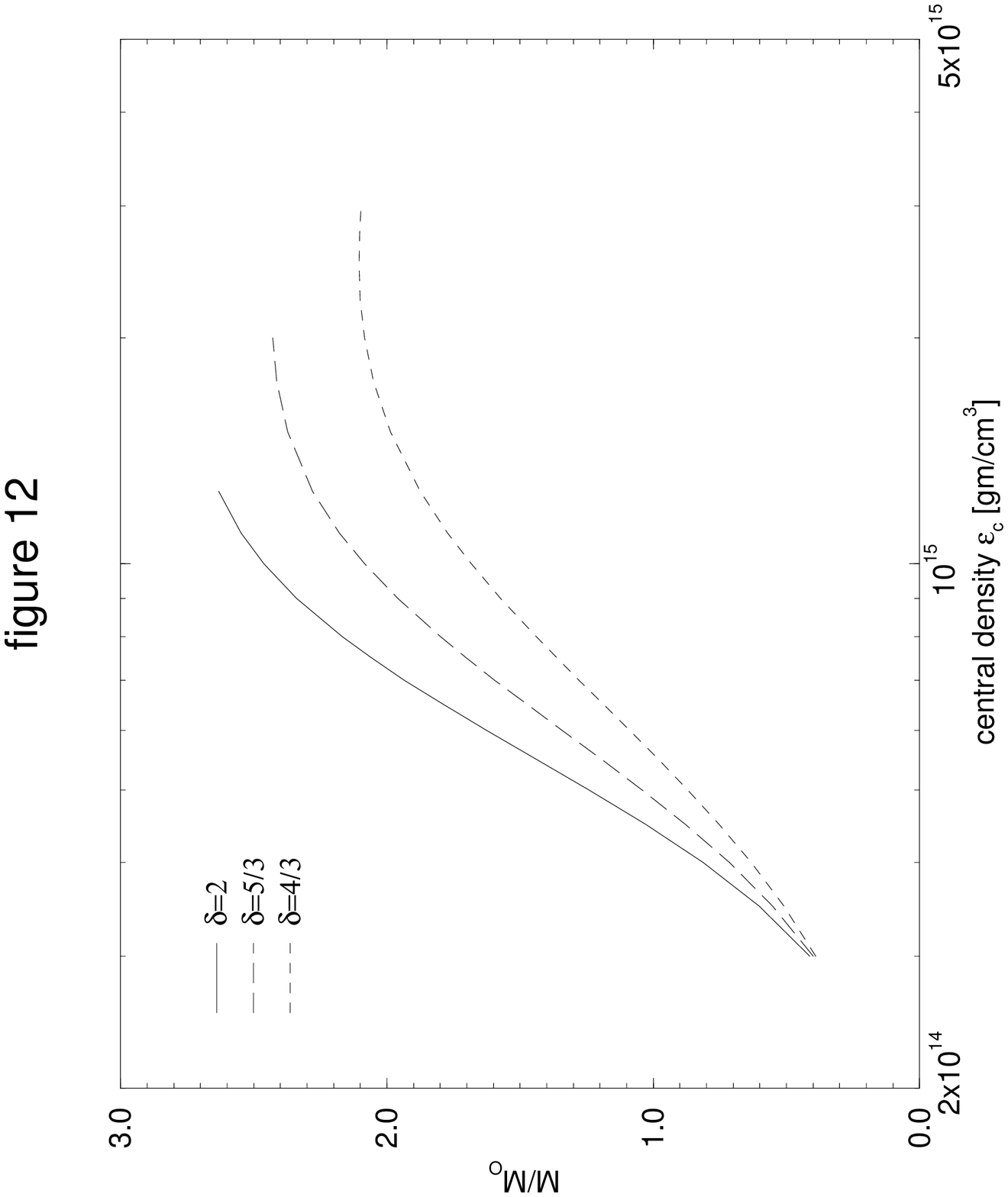}
\newpage
\epsffile{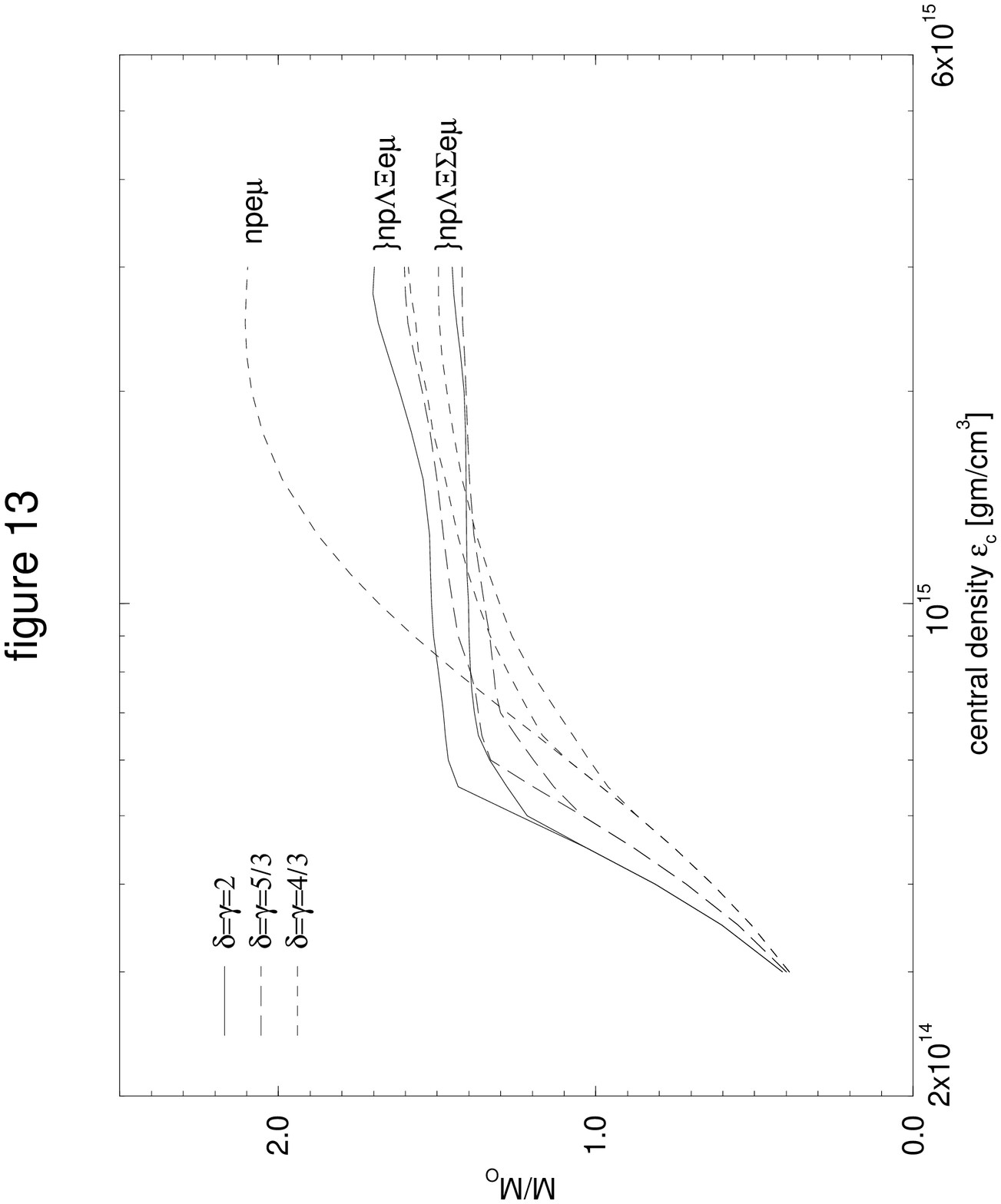}

\end{document}